\UseRawInputEncoding
\documentclass[a4paper,11pt]{article}
\pdfoutput=1 % if your are submitting a pdflatex (i.e. if you have
\usepackage{jheppub} % for details on the use of the package, please
\usepackage[T1]{fontenc} % if needed
\usepackage{bm}
\usepackage{mathrsfs}
\usepackage{amsbsy}
\usepackage{amssymb}
\usepackage{graphicx}
\usepackage{subfigure}
\usepackage{amsmath}
\usepackage{amsfonts}
\usepackage{xcolor}

\newcommand{\be}{\begin{equation}}
\newcommand{\ee}{\end{equation}}
\newcommand{\bea}{\begin{eqnarray}}
\newcommand{\eea}{\end{eqnarray}}

\title{\boldmath Pole-skipping points in 2D gravity and SYK model}

\author[a]{Haiming Yuan,}
\author[a,1]{Xian-Hui Ge,\note{Corresponding author.}}
\author[b,c]{Keun-Young Kim,}
\author[b]{Chang-Woo Ji,}
\author[d,e]{and Yongjun Ahn}

\affiliation[a]{Department of Physics, College of Sciences, Shanghai University,\\99 Shangda Road, 200444 Shanghai, China}
\affiliation[b]{Department of Physics and Photon Science, Gwangju Institute of Science and Technology,\\123 Cheomdan-gwagiro, Gwangju 61005, Korea}
\affiliation[c]{Research Center for Photon Science Technology, Gwangju Institute of Science and Technology,\\123 Cheomdan-gwagiro, Gwangju 61005, Korea}
\affiliation[d]{Wilczek Quantum Center, School of Physics and Astronomy,
Shanghai Jiao Tong University, Shanghai 200240, China}
\affiliation[e]{Shanghai Research Center for Quantum Sciences, Shanghai 201315, China}
\emailAdd{yuan\_haiming@shu.edu.cn}
\emailAdd{gexh@shu.edu.cn}
\emailAdd{fortoe@gist.ac.kr}
\emailAdd{physianoji@gm.gist.ac.kr}
\emailAdd{yongjunahn619@gmail.com}

\abstract{
We represent the first investigation of pole-skipping on both the gravity and field theory sides. In contrast to the higher dimensional models, there is no momentum degree of freedom in $(1+1)-$dimensional bulk theory. Thus, we then consider a scalar field mass as our degree of freedom for the pole-skipping phenomenon instead of momentum. The pole-skipping frequencies of the scalar field in 2D gravity are the same as higher dimensional cases: $\omega=-i2\pi Tn$ for positive integers $n$. At each of these frequencies, there is a corresponding pole-skipping mass, so the pole-skipping points exist in $(\omega,m)$ space. We also compute the pole-skipping points of the SYK model in $(\omega, h)$ space where $h$ is the dimension of the bilinear primary operator. We find that there is a one-to-one correspondence of the pole-skipping points between the JT gravity and the SYK model. To obtain the pole-skipping points, we need to consider the parameter  $\epsilon$ related to the chemical potential on the horizon of charged JT gravity and the particle-hole asymmetric parameter $\mathcal{E}$ of the complex SYK model as shift parameters. This highlights the $\epsilon-\mathcal{E}$ correspondence in relation to pole-skipping phenomenon. }

\begin{document}
\maketitle
\flushbottom

\section{Introduction} \label{sec:intro}
\quad The AdS/CFT correspondence~\cite{maldacena,witten1,witten2,gubser} provides a useful method to compute Green's functions in strongly coupled quantum many-body systems~\cite{casalderrey,natsuume,ammon,zaanen,hartnoll}. In recent years, pole-skipping, an interesting property that occurs in the Green's function, has been studied in much of the literature~\cite{Grozdanov1,Blake1,Grozdanov2,Makoto1,Makoto2,BlakeDavison,Das,Abbasi2,Abbasi1,Choi,Karunava,Yongjun2,Mahdi}. The pole-skipping phenomenon is a mathematical property that the retarded Green's function is not unique at some points in the complex momentum space $(\omega,k)$. From now on, we call those points pole-skipping points. Generally, the retarded Green's function takes a form
\be
G^R(\omega,k)=\frac{b(\omega,k)}{a(\omega,k)}\,.
\ee
At a pole-skipping point $(\omega_\star, k_\star)$, both $a$ and $b$ satisfy $a(\omega_\star, k_\star)=b(\omega_\star, k_\star)=0$, and the retarded Green's function becomes ill-defined. So, if we find the intersections of zeros and poles in the retarded Green's functions, we can obtain pole-skipping points.
%\textcolor{blue}{(Is 3/2 spin field not related to the quantum chaos?)} \textcolor{orange}{[CW: Suggestion of little relation between chaos and pole-skipping for 3/2 spin fields has been commented in \cite{Ceplak:2021efc}].}
%There is a scalar field coupled to gravity, whose mass is equal to the Breitenlohner-Freedman bound of AdS$_2$ \cite{Das2}.
For the theories with the holographic duals, we can use a simpler method to obtain the pole-skipping points from the bulk field equation \cite{Makoto1,Makoto2,BlakeDavison}. The absence of a unique incoming mode at the black hole horizon leads to the nonuniqueness of the Green's function on the boundary. For the static black holes, the leading pole-skipping frequency $\omega$ is given by \cite{Makoto3,N1,Yuan1,N2,Yuan2,Diandian}
\bea
\label{eq:43}
\omega=2\pi i T(s-1),
\eea
where $i$ is the imaginary unit, and $s$ denotes the spin of the operator. For instance, it has been noted that pole-skipping points in the upper-half $\omega$-plane are linked to quantum chaos and are closely related to hydrodynamics. In the maximally chaotic systems, the Lyapunov exponent $\lambda$ and the butterfly velocity $v_B$ can be extracted from the leading pole-skipping point of the metric fluctuation ($s=2$) \cite{Grozdanov1,Blake1,Blake3,Jeong}. For other fluctuations ($s=0,1/2,1,3/2$), the leading pole-skipping points are not related to the maximal chaos~\cite{N2}.

The Jackiw-Teitelboim (JT) gravity is a two-dimensional theory of gravity coupled to a dilaton, and describes the near horizon dynamics of near extremal black holes~\cite{Jackiw,Teitelboim}. The Sachdev-Ye-Kitaev (SYK) model is a strongly interacting quantum system at low energy. It has some interesting properties: It is solvable at the large $N$ limit with an emergent conformal symmetry~\cite{Sachdev,Kitaev}. The SYK model is maximally chaotic so that the Lyapunov exponent saturates the Maldacena-Shenker-Stanford(MSS) bound $\lambda=2\pi T$~\cite{Maldacena:2015waa,Jensen}. Symmetries and reparametrization connect the SYK model to the AdS$_2$ horizon. Near the horizon of an AdS$_2$ black hole corresponds to the IR of Majorana SYK~\cite{Maldacena}. JT gravity has a nearly AdS$_2$ geometry, and its conformal symmetry is slightly broken. The Schwarzian action can be derived for JT gravity by spontaneously breaking the reparametrization symmetry down to $SL(2,\mathbb{R})$~\cite{Sarosi}. Similarly, the Majorana SYK model also has its conformal symmetry spontaneously broken to $SL(2,\mathbb{R})$ due to reparametrization, allowing for the calculation of the Schwarzian action~\cite{Stanford,Polchinski}. The effective action can be obtained from both gravity theory on the boundary and the Majorana SYK model. Therefore, JT gravity has been considered as the gravitational dual of the Majorana SYK model. In the bulk gravity models, there are also various analogies to the complex SYK model \cite{csyk1,csyk2}. For example, for nearly AdS$_2$ in JT gravity with $SL(2,L)\times U(1)$ symmetry, it is holographically dual to the complex SYK model \cite{Godet}.

Pole-skipping has primarily been studied on the gravity side thus far~\cite{Grozdanov1,Blake1,Grozdanov2,Makoto1,Makoto2,BlakeDavison}. In the case of the (1+1)-dimensional SYK chain, it has been discovered that pole-skipping is determined by the stress tensor's contribution to many-body chaos, even when it is not at its maximum~\cite{Choi}. However, up to now, there are few works that reported the pole-skipping points on both the gravity and field theory sides in previous literature. There have been no findings regarding the connection between pole-skipping points in the SYK model and those in JT gravity.

In this paper, we aim to fill this gap. Note that the Green's functions of the JT gravity and SYK model do not depend on momentum $k$ because there is no spatial coordinate as they are $(0+1)-$dimensional field theories. %We show that the pole-skipping points can not be computed with only the frequency $\omega$. %\textcolor{red}{(The meaning of the following is not clear. If it does not have enough meaning, we may remove it.) For the massive Klein-Gordon equation, the dispersion relation yields $-p_\mu p_\mu=\omega^2-k^2=m^2$.}
In the absence of momentum $k$, one way to define the pole-skipping points is to identify the mass $m$ as a variable of the Green's function.
Following Ref.~\cite{Ahn:2020bks}, we compute the pole-skipping points in the $(\omega, m)$ space in three ways: 1) near horizon analysis 2) analytic bulk solution for the JT gravity, and 3) conformal two-point function of the SYK model. We have obtained consistent results by all three methods and found a one-to-one correspondence of the pole-skipping between the JT gravity and the SYK model.
%We first investigate the pole-skipping points in 2D gravity in two different ways: 1) near horizon analysis and 2) analytic bulk solution.
%{\color{magenta}We find that the frequencies of pole-skipping points of scalar field and Fermionic field in 2D gravity from point 1) are the same as the results of higher dimension. However, the number of the frequencies of pole-skipping points obtained by point 2) is more abundant than that obtained by point 1). We then move to point 3), and the results of point 3) are consistent with those of point 2). We find one-to-one correspondence of pole-skipping between the JT gravity and the SYK model.}
%In the aspect of scaling dimension, we will show that $\Delta_+$ and $\Delta_-$ in bulk scalar field correspond to $h$ and $h_-$ in SYK model.

The structure of this paper is as follows: In Sec.~\ref{sec:2}, we calculate the pole-skipping points in the JT gravity and the Majorana SYK model. In the JT gravity, we introduce a massive scalar probe and calculate the pole-skipping points of a dual operator. We also study pole-skipping in the Majorana SYK models. The results show good agreement with each other.
%\textcolor{red}{(The following is too much detail to put in the introduction. We propose to remove it. Otherwise we have to explain all notations here.)For dimension $h$ and our newly defined dimension $h_-$, we will show the pole-skipping points in space $(\omega'',h)$ and $(\omega'',h_-)$ in Majorana SYK model are the same as the pole-skipping points in space $(\omega,\Delta_+)$ and $(\omega,\Delta_-)$ in bulk scalar retarded Green's function.}
In Sec.~\ref{sec:3}, we compare the pole-skipping points in charged JT gravity with those in the complex SYK model. We show that the electric field strength and the spectral asymmetry shift the pole-skipping frequencies in each model. The correspondence between the 2D bulk scalar field and the 1D SYK model is a good example of holographic duality. We summarize and conclude in Sec.~\ref{sec:Discussion}
\section{Comparison of pole-skipping in JT gravity and Majorana SYK model}
\label{sec:2}
\subsection{Pole-skipping in JT gravity}
\label{sec:Scalar}
\quad As mentioned earlier, pole-skipping can be seen in Green's function. One can compute a retarded Green's function of an operator dual to a bulk field from the solution to the corresponding classical equation of motion. Our interest is the pole-skipping of the retarded Green's function of the scalar operator $\mathcal{O}$, which lives in the field theory dual to JT gravity.

The action of JT gravity is given by~\cite{Jackiw,Teitelboim}
\bea
\label{eq:41}
S=\frac{1}{2G}\int d^2x\sqrt{-g}\phi\left(R+\frac{2}{L^2}\right).
\eea
One can obtain a  Schwarzschild-like solution from the action \eqref{eq:41} with a gauge choice $\phi=\frac{r}{L}$ \cite{Louis-Martinez,Achucarro,Gegenberg}
\bea
\label{eq:1}
ds^2=-f(r)dt^2+\frac{1}{f(r)}dr^2,
\eea
where
\bea
f(r)=r^2/L^2-2GLM,
\eea
where $M$ is the mass of the black hole. This is an asymptotic AdS$_2$ metric with the event horizon at
\bea
r_h=L\sqrt{2GLM}.
\eea
For convenience, we set $2GLM=1$ in the following. The Hawking radiation temperature is given by
\bea
T=\frac{f'(r)}{4\pi}\bigg|_{r=r_h}=\frac{1}{2\pi L}.
\eea
\subsubsection{Pole-skipping: near-horizon analysis}
\label{sec:ScalarNear}
\quad The near horizon analysis provides us a shortcut to calculate pole-skipping points of a holographic retarded Green's function. In this subsection, we calculate pole-skipping points by analyzing the near-horizon behavior of a bulk field. We consider the minimally coupled massive scalar field $\Psi$ of which the dynamics are given by the Klein-Gordon equation as
\bea
\label{eq:3}
\frac{1}{\sqrt{-g}}\partial_\mu(\sqrt{-g}g^{\mu\nu}\partial_\nu\Psi)-m^2\Psi=0\,,
\eea
where $m$ is the mass of the scalar field. For our convenience, we introduce the incoming Eddington-Finkelstein(EF) coordinate $v=t+r_*$, where $r_*$ is the tortoise coordinate $dr_*:=dr/f(r)$. In the incoming EF coordinate,~\eqref{eq:1} becomes
\bea
\label{eq:2}
ds^2=-f(r)dv^2+2dvdr,
\eea
and the Klein-Gordon~\eqref{eq:3} equation can be expanded as
\bea
\label{eq:4}
f(r)\Psi''(r)+(f'(r)-2i\omega)\Psi'(r)-m^2\Psi(r)=0,
\eea
where we take Fourier transformation as $\Psi(v,r)=e^{-i \omega v}\Psi(r)$. From now on we will use scaled frequency $\mathfrak{w}=\frac{\omega}{2\pi T}$, and mass $\mathfrak{m}=\frac{m}{2\pi T}$.

In holographic models, the retarded Green's function is encoded in the classical solution which satisfies the incoming boundary condition at the horizon. Since the blackening factor goes $f(r) = 4\pi T(r-r_h) + \mathcal{O}\left(r-r_h\right)^2$ near the horizon, one can check that the Klein-Gordon equation~\eqref{eq:4} has a regular singularity at $r=r_h$. The singularity can be seen if we approximate \eqref{eq:4} to
\bea
\label{eq:approxhor}
\Psi''+\frac{1-i\mathfrak{w}}{r-r_h}\Psi'-\frac{\pi T\mathfrak{m}^2}{r-r_h}\Psi=0
\eea
near horizon. According to conventional differential equation techniques, one can solve \eqref{eq:4} by imposing series solution ansatz as
\bea
\label{eq:seransatz}
\Psi(r)=(r-r_h)^\lambda\sum^\infty_{p=0}\Psi_p(r-r_h)^p.
\eea
At the lowest order, we can obtain the indicial equation $\lambda(\lambda-i\mathfrak{w})=0$. The two roots are
\bea
\label{eq:7}
\lambda_1=0,\quad \lambda_2=i\mathfrak{w}\,.
\eea

Generally, a solution with exponent $\lambda_1$ is regular at the horizon. Therefore, we can obtain a unique `ingoing' solution with exponent $\lambda_1$, if $\lambda_2$ is not an integer. However, if $\lambda_2$ is an integer, there might be an additional regular solution which is a signal of pole-skipping at the horizon. For example, let us consider $i\mathfrak{w}=1$. According to a standard technique for a differential equation, the following solution ansatz is more appropriate than \eqref{eq:seransatz}:
\bea
\Psi(r)=\sum^\infty_{p=0}\Psi_{1,p}(r-r_h)^p + (r-r_h)\log(r-r_h)\sum^\infty_{q=0}\Psi_{2,q}(r-r_h)^q\,,
\eea
where we include $\log$ term. After substituting it into the Klein-Gordon eqaution~\eqref{eq:4}, up to $\mathcal{O}(r-r_h)^0$ the equation of motion becomes
\bea
(1-i\mathfrak{w})\left(\Psi_{1,1}+ \Psi_{2,0}\log(r-r_h)\right)-\left(\pi T \mathfrak{m}^2\Psi_{1,0}-(2-i \mathfrak{w})\Psi_{2,0}\right)=0\,.
\eea
The first term vanishes at $i\mathfrak{w}=1$. If we focus on $\mathfrak{m}=0$, we obtain $\Psi_{2,0}=0$, and the series solution takes two independent regular solutions as follows:
\bea
\Psi(r)=\Psi_{1,0} + (r-r_h)^1\sum^\infty_{p=0}\tilde{\Psi}_{2,p}(r-r_h)^p\,,
\eea
where two independent coefficients are $\Psi_{1,0}$ and $\tilde{\Psi}_{2,0}$. Thus the first pole-skipping point is
\bea
\label{eq:8}
\mathfrak{w}_{\star}=-i,\quad \mathfrak{m}^2_{\star}=0\,.
\eea

In general, at the pole-skipping points, the series solution takes the following form:
\bea
\Psi(r)=\sum^{n-1}_{p=0}\Psi_{1,p}(r-r_h)^p + (r-r_h)^n\sum^\infty_{q=0}\tilde{\Psi}_{2,q}(r-r_h)^q\,,
\eea
where $\Psi_{1,p}$ and $\tilde{\Psi}_{2,q}$ are independent coefficients.
There is a systematic procedure to calculate pole-skipping points~\cite{BlakeDavison}. Firstly, we expand $\Psi(r)$ with a Taylor series
\bea
\label{eq:10}
\Psi(r)=\sum^\infty_{p=0}\Psi_p(r-r_h)^p=\Psi_0+\Psi_1(r-r_h)+\Psi_2(r-r_h)^2+\dots.
\eea
We substitute \eqref{eq:10} into \eqref{eq:4} and expand the equation of motion in powers of $(r-r_h)$. Then, a series of the perturbed equation in the order of $(r-r_h)$ can be denoted as
\bea
\label{eq:11}
S=\sum^\infty_{p=0}S_p(r-r_h)^p=S_0+S_1(r-r_h)+S_2(r-r_h)^2+\cdots\,.
\eea
We write down the first few equations $S_p=0$ in the expansion of \eqref{eq:11}
\begin{equation}
\begin{split}
&0=M_{11}(\omega,m)\Psi_0+(2\pi T-i\omega)\Psi_1,\\
&0=M_{21}(\omega,m)\Psi_0+M_{22}(\omega,m)\Psi_1+(4\pi T-i\omega)\Psi_2,\\
&0=M_{31}(\omega,m)\Psi_0+M_{32}(\omega,m)\Psi_1+M_{33}(\omega,m)\Psi_2+(6\pi T-i\omega)\Psi_3\,.
\end{split}
\end{equation}
To find an incoming solution, we should solve a set of linear equations of the form
\bea
\label{eq:12}
\mathcal{M}(\omega,m)\cdot \Psi\equiv\left(\begin{array}{ccccc}
    M_{11} & (2\pi T-i\omega) & 0    & 0  &\dots\\
    M_{21} & M_{22}& (4\pi T-i\omega)& 0   &\dots\\
    M_{31} & M_{32}&  M_{33} &(6\pi T-i\omega) &\dots\\
    \dots   &  \dots&  \dots  &\dots   &\dots\\
\end{array}\right)\left(\begin{array}{ccccc}
   \Psi_0\\
   \Psi_1\\
    \Psi_3 \\
    \dots \\
\end{array}\right)=0\,.
\eea
The $n$-th pole-skipping points $(\omega_{n}, m_{n})$ can be calculated by solving
\bea
\label{eq:42}
\omega_{\star n}=-i2\pi Tn,\qquad {\rm det}\mathcal{M}^{(n)}(\omega_\star,m_\star)=0\,,
\eea
where the matrix $\mathcal{M}^{(n)}$ is the $(n\times n)$ square matrix whose elements are taken from $M_{11}$ to $M_{nn}$ in \eqref{eq:12}.\footnote{The first few elements of this matrix have been shown in Appendix~\ref{sec:Details1}.}
The following are the resulting pole-skipping points:
\begin{equation}
\label{eq:pspresult1}
\begin{split}
\mathfrak{w}&=-i, \quad \mathfrak{m}^2=0\,;\\
\mathfrak{w}&=-2i, \quad \mathfrak{m}^2=0,2\,;\\
\mathfrak{w}&=-3i, \quad \mathfrak{m}^2=0,2,6\,;\\
&\qquad \qquad \vdots
\end{split}
\end{equation}

Pole-skipping points~\eqref{eq:pspresult1} can be rewritten in terms of operator dimension. Note that the boundary asymptotic behavior of $\Psi$ takes the following form
\begin{equation}
\label{eq:boundaryasymp}
\begin{split}
\Psi(r) &= A(\omega, m)r^{-\Delta_{-}}+B(\omega, m)r^{-\Delta_+}+\cdots\,,
\end{split}
\end{equation}
where the conformal dimensions $\Delta_\pm=\frac{1}{2}\pm\frac{1}{2}\sqrt{1+4m^2L^2}.$
For standard quantization, we identify the operator dimension as $\Delta_+$. Pole-skipping points are
\begin{equation}
\label{eq:pspresultstd}
\begin{split}
\mathfrak{w}&=-i, \quad \Delta_+=1\,;\\
\mathfrak{w}&=-2i, \quad \Delta_+=1,2\,;\\
\mathfrak{w}&=-3i, \quad \Delta_+=1,2,3\,;\\
&\qquad \qquad \vdots
\end{split}
\end{equation}
For alternative quantization, we identify the operator dimension as $\Delta_-$. Pole-skipping points are
\begin{equation}
\label{eq:pspresultalt}
\begin{split}
\mathfrak{w}&=-i, \quad \Delta_-=0\,;\\
\mathfrak{w}&=-2i, \quad \Delta_-=0,-1\,;\\
\mathfrak{w}&=-3i, \quad \Delta_-=0,-1,-2\,;\\
&\qquad \qquad \vdots
\end{split}
\end{equation}
%At $n$-th pole-skipping points, any value of $\Psi_0$ and $\Psi_n$ satisfies the equation \eqref{eq:4}; there are two independent free parameters $\Psi_0$ and $\Psi_n$ in the general series solution \eqref{eq:seransatz} to this equation.

\subsubsection{Pole-skipping from Green's function}
\label{sec:ScalarBoundary}
\quad In this subsection, we calculate the retarded Green's function on the boundary to obtain pole-skipping points. To solve fluctuation equation \eqref{eq:3} analytically, it is convenient to introduce coordinate $u$, which is defined as $r=L {\rm cosh}(u)$. In the $u$ coordinate, metric \eqref{eq:1} takes following form
\bea
\label{eq:14}
ds^2=-{\rm sinh}^2(u)dt^2+L^2du^2,
\eea
and we can expand the Klein-Gordon equation \eqref{eq:3} as
\bea
\label{eq:15}
\Psi''(u)+{\rm coth}(u)\Psi'(u)+L^2(-m^2+\omega^2{\rm csch}^2(u))\Psi(u)=0.
\eea
After taking the change of variable $z={\rm tanh}^2(u)$, we use the ansatz
\bea
\Psi(u)=(1-z)^{\Delta_+/2}z^{-i\omega L/2}F(z)
\eea
to simplify the above equation. Eq.~\eqref{eq:15} become
\bea
\label{eq:65}
&&z(1-z)\frac{d^2F(z)}{dz^2}+\bigg(1-2z-\frac{z}{2}\sqrt{1+4m^2L^2}+i\omega L(z-1)\bigg)\frac{dF(z)}{dz}\nonumber\\
&&-\frac{1}{4}\bigg(1+\sqrt{1+4m^2L^2}-i\omega L(2+\sqrt{1+4m^2L^2})-L^2(\omega^2-m^2)\bigg)F(z)=0.
\eea
We find Eq.~\eqref{eq:65} fits into the standard hypergeometric differential equation
\be
\label{eq:16}
z(1-z)\frac{d^2F(z)}{dz^2}+(c-(1+a+b)z)\frac{dF(z)}{dz}-abF(z)=0,
\ee
where
\begin{equation}
\label{Eq:hyper2}
\begin{split}
a&=\frac{1}{2}(3+\sqrt{1+4m^2L^2}-2i\omega L)\,,\\
b&=\frac{1}{2}(1-\sqrt{1+4m^2L^2}-2i\omega L)\,,\\
c&=1-i\omega L\,.
\end{split}
\end{equation} The horizon is located at $z=0$ while the boundary is located at $z=1$. The incoming solution is given by
\bea
\label{eq:32}
F_{in}=\,_2F_1(a,b,a+b-{\cal N};z),
\eea
where we define ${\cal N}=\sqrt{\frac{1}{4}+m^2L^2}$. Near the boundary $(z \rightarrow 1)$, the asymptotic behavior of \eqref{eq:32} takes following form
\bea
\label{eq:33}
F_{in}\approx (1-z)^{-{\cal N}}A(\omega,m)+B(\omega,m)+C(\omega,m){\rm log}(1-z).
\eea
The conformal dimensions can be written as $\Delta_{\pm}=\frac{1}{2} \pm \cal N$.

If ${\cal N}$ are non-integers,\footnote{If ${\cal N}$ are integers the Green's function is different from \eqref{eq:17}. We do not consider this case here because it is irrelevant to the pole-skipping points originating from the near-horizon dynamics.} $C(\omega,m)$ in Eq.~\eqref{eq:33} vanishes. For the standard quantization, $A(\omega, m)$ is the source and $B(\omega, m)$ is the vacuum expectation value for the operator $\mathcal{O}$ of dimension $\Delta_{+}$. The corresponding retarded Green's function is
\bea
\label{eq:17}
G^R_{\mathcal{O}\mathcal{O}}&\propto&\frac{B}{A}\nonumber\\
&\propto&(2\pi T)^{2\Delta_+-1}\frac{\Gamma\big(\frac{1}{4}(3+\sqrt{1+4m^2L^2}-2i\omega L)\big)\Gamma\big(\frac{1}{4}(1+\sqrt{1+4m^2L^2}-2i\omega L)\big)}{\Gamma\big(\frac{1}{4}(3-\sqrt{1+m^2L^2}-2i\omega L)\big)\Gamma\big(\frac{1}{4}(1-\sqrt{1+m^2L^2}-2i\omega L)\big)}\frac{\Gamma(\frac{1}{2}-\Delta_{+})}{\Gamma(-\frac{1}{2}+\Delta_{+})}\nonumber\\
&=&(2\pi T)^{2\Delta_+-1}\frac{\Gamma(\frac{1}{2}+\frac{\Delta_{+}}{2}-\frac{iL\omega}{2})\Gamma(\frac{\Delta_{+}}{2}-\frac{iL\omega}{2})}{\Gamma(1-\frac{\Delta_{+}}{2}-\frac{iL\omega}{2})\Gamma(\frac{1}{2}-\frac{\Delta_{+}}{2}-\frac{iL\omega}{2})}
\frac{\Gamma(\frac{1}{2}-\Delta_{+})}{\Gamma(-\frac{1}{2}+\Delta_{+})}\nonumber\\
&=&(\pi T)^{2\Delta_+-1}\frac{\Gamma(\Delta_{+}-\frac{i\omega}{2\pi T})}{\Gamma(1-\Delta_{+}-\frac{i\omega}{2\pi T})}\frac{\Gamma(\frac{1}{2}-\Delta_{+})}{\Gamma(-\frac{1}{2}+\Delta_{+})}\,.
\eea
We have used the Legendre duplication formula $\Gamma(z)\Gamma(z+\frac{1}{2})=2^{1-2z}\pi^{1/2}\Gamma(2z)$ above.

Since $\Gamma(-n)$  diverges for non-negative integer $n$, the lines of pole and zero of the Green's function~\eqref{eq:17} are
\bea
\label{eq:34}
\left\{
\begin{aligned}
\Delta_{+}-\frac{i\omega}{2\pi T}&=&0,-1,-2,-3,\cdots\,, \qquad\text{(lines of pole)}\,,\\
1-\Delta_{+}-\frac{i\omega}{2\pi T}&=&0,-1,-2,-3,\cdots\,,\qquad\text{(lines of zero)}\,.
\end{aligned}
\right.
\eea
Graphically a pole-skipping point is an intersection of a line of poles and a line of zeros. The pole-skipping points for~\eqref{eq:34} are
\begin{equation}
\label{eq:pspJT}
\begin{split}
%\mathfrak{w}&=-\frac{i}{2}, \quad \Delta_{+}=\frac{1}{2}\,,\\
\mathfrak{w}&=-i, \quad \Delta_{+}=1\,,\\
%\mathfrak{w}&=-\frac{3i}{2}, \quad \Delta_{+}=\frac{1}{2},\frac{3}{2}\,,\\
\mathfrak{w}&=-2i, \quad  \Delta_{+}=1,2\,,\\
%\mathfrak{w}&=-\frac{5i}{2}, \quad \Delta_{+}=\frac{1}{2},\frac{3}{2},\frac{5}{2}\,,\\
\mathfrak{w}&=-3i, \quad \Delta_{+}=1,2,3\,,\\
%\mathfrak{w}&=-\frac{7i}{2}, \quad \Delta_{+}=\frac{1}{2},\frac{3}{2},\frac{5}{2},\frac{7}{2}\,,\\
%\mathfrak{w}&=-4i, \quad \Delta_{+}=1,2,3,4\,,\\
&\qquad \qquad \vdots
\end{split}
\end{equation}
which agrees with the pole-skipping points obtained from near-horizon behavior.

We can also calculate pole-skipping points for alternative quantization in $(\omega, \Delta_-)$ space. For alternative quantization, $A(\omega,m)$ is the vacuum expectation value and $B(\omega, m)$ is the source in \eqref{eq:33}. The retarded Green's function is an inverse of \eqref{eq:17}:
\begin{equation}
\label{eq:GRalt}
\begin{split}
G^R_{\mathcal{O}\mathcal{O}}&\propto
\frac{A}{B}\\
&\propto (2\pi T)^{-2{\cal N}}\frac{\Gamma\big(\frac{1}{4}(3-\sqrt{1+m^2L^2}-2i\omega L)\big)\Gamma\big(\frac{1}{4}(1-\sqrt{1+m^2L^2}-2i\omega L)\big)\Gamma(-{\cal N})}{\Gamma\big(\frac{1}{4}(3+\sqrt{1+4m^2L^2}-2i\omega L)\big)\Gamma\big(\frac{1}{4}(1+\sqrt{1+4m^2L^2}-2i\omega L)\big)\Gamma({\cal N})}\,,
\end{split}
\end{equation}
where we take the first line of \eqref{eq:17}. Note that in terms of $m$, the pole-skipping points of \eqref{eq:GRalt} are the same as those of \eqref{eq:17}. As we have already seen in \eqref{eq:pspresultalt}, in terms of operator dimension $\Delta_-$, the pole-skipping points of the alternative quantization are
\begin{equation}
\label{eq:pspJTalt2}
\begin{split}
\mathfrak{w}&=-i, \quad \Delta_-=0\,;\\
%\mathfrak{w}&=-\frac{3i}{2}, \quad \Delta_-=-\frac{1}{2}\,;\\
\mathfrak{w}&=-2i, \quad  \Delta_-=0,-1\,;\\
%\mathfrak{w}&=-\frac{5i}{2}, \quad \Delta_-=-\frac{1}{2},-\frac{3}{2}\,;\\
\mathfrak{w}&=-3i, \quad \Delta_-=0,-1,-2\,;\\
%\mathfrak{w}&=-\frac{7i}{2}, \quad \Delta_-=-\frac{1}{2},-\frac{3}{2},-\frac{5}{2}\,;\\
&\qquad \qquad \vdots
\end{split}
\end{equation}

We also calculated the pole-skipping points of Dirac field in Appendix~\ref{sec:Dirac}. The locations of the pole-skipping points from near horizon analysis correspond to what we obtained from bulk Green's function on the boundary. This can serve as a reference for pole-skipping of fermionic field in two-dimensional gravity.
\subsection{Pole-skipping in Majorana SYK model}
\label{sec:SYK}
\quad The Majorana Sachdev-Ye-Kitaev (SYK) model describes $N$ Majorana fermions with random $q-$body interactions, which is solvable in the large $N$ and low energy limit \cite{Sachdev,Maldacena,Kitaev}. The SYK model shows emergent conformal symmetry in the low-temperature limit, and the breaking of the conformal symmetry leads to a finite Schwarzian action for reparametrization modes \cite{Maldacena}. This Schwarzian action is the same as the boundary action of the Jackiw-Teitelboim gravity in nearly AdS$_2$ spacetime \cite{Kitaev}. The Hamiltonian is given as \cite{Stanford}
\bea
H=(i)^{q/2}\sum_{1\leq x_1<x_2\dots<x_q\leq N}j_{x_1x_2\dots x_q}f_{x_1}f_{x_2}\dots f_{x_q},
\eea
where $i$ is the imaginary unit, and $f_i$ is a fermion obeying the anticommutation relations. The $j_{x_1x_2\dots x_q}$ are independent Gaussian random couplings with zero mean that satisfy
\bea
\langle j^2_{x_1x_2\dots x_q}\rangle=\frac{J^2(q-1)!}{N^{q-1}},
\eea
where dimension-one parameter $J$ is the energy scale. This model is conformal symmetry in the infrared limit, and the effective action is ~\cite{Gross1,Kitaev2,Rosenhaus,Maria}
\bea
S_{\rm CFT}=-\frac{N}{2}{\rm log}\;{\rm det}(-\Sigma)+\frac{N}{2}\int d\tau_1d\tau_2\bigg(\Sigma(\tau_1,\tau_2)G(\tau_1,\tau_2)-\frac{J^2}{q}\vert G(\tau_1,\tau_2)\vert^q\bigg)
\eea
$S_{\rm CFT}$ is invariant under the reparametrization $\tau\rightarrow f(\tau)$ \cite{Kitaev,Stanford}
\bea
G(\tau_1,\tau_2)\rightarrow f'(\tau_1)^\Delta f'(\tau_2)^\Delta G(f(\tau_1),f(\tau_2))
\eea
provided that $\Delta=\frac{1}{q}$. The saddle of $S_{\rm CFT}$ can solve for a conformal field theory (CFT) two-point function
\bea
G^c(\tau_1,\tau_2)=-\frac{1}{N}\langle f_i(\tau_1)f_i(\tau_2)\rangle=-b\frac{{\rm sgn}(\tau_{12})}{\vert \tau_{12}\vert^{2\Delta}},
\eea
where $b$ is a dimensionless constant. The expression for the full two-point function is
\bea
G(\tau_1,\tau_2)=-\frac{1}{Z}\int D f \frac{1}{N}f_i(\tau_1)f_i(\tau_2)e^{-S_{\rm SYK}}.
\eea
In interacting field theory, it is natural to use the conformal perturbation method to express action and Green's function. To compute a higher point correlation function, one should view it as leading order and move away from the infrared limit. The Green's function can be expanded as
\bea
G(\tau_1,\tau_2)=&&-\frac{1}{N}\langle f_i(\tau_1)f_i(\tau_2)\rangle+\sum_h g_h\int d\tau_3\;\frac{1}{N}\langle f_i(\tau_1)f_i(\tau_2)\mathcal{O}_h(\tau_3)\rangle\nonumber\\
&&-\frac{1}{2}\sum_{h,h'} g_h g_{h'}\int d\tau_3d\tau_4\;\frac{1}{N}\langle f_i(\tau_1)f_i(\tau_2)\mathcal{O}_h(\tau_3)\mathcal{O}_{h'}(\tau_4)\rangle+\dots.
\eea
At strong coupling, a tower of fermionic bilinear, primary, single-trace operator forms $\mathcal{O}_{h}=\sum^N_{i=1}\chi_i\partial^{h}_\tau\chi_i$ \cite{Gross3,Gross2}. The action can be described as a CFT fixed point deformed by an infinite set of irrelevant primary operators
\bea
\label{eq:actionMSYKeq:pspresult1}
S_{\rm SYK}=S_{\rm CFT}+\sum_h g_h\int d\tau\; \langle \mathcal{O}_h(\tau)\rangle-\frac{1}{2}\sum_{h,h'} g_h g_{h'}\int d\tau_1d\tau_2\;\langle \mathcal{O}_h(\tau_1)\mathcal{O}_{h'}(\tau_2)\rangle+\dots,\nonumber
\eea
where $\mathcal{O}_h(\tau)$ is a bilinear, primary, and single-trace operator which has scaling dimension $h$.
%\textcolor{red}{Denote the perturbed action
%\bea
%\delta S=\sum_h g_h\int d\tau\; \mathcal{O}_h(\tau).
%\eea
%[YJ: Is this necessary?]}
Using linear response theory, one often adds an external source and sees the response to the operator $\mathcal{O}_h$, $\delta\langle \mathcal{O}_h\rangle$, which couples to the external source $g_h$. We regard $\mathcal{O}_h$ as scalar operators of boundary theory and hope that it can be dual to scalar fields $\Psi$ of bulk theory through AdS/CFT dictionary in our near-horizon analysis.

In this section, we will show that the pole-skipping points in the Majorana SYK model are the same as those in the JT gravity. The retarded Green's function in frequency space can be expanded as \cite{Maria}
\bea
G_R(\omega)=G_R^c(\omega)+\delta G_R(\omega).
\eea
The conformal part is given by
\bea
G_R^c(\omega)=-i\frac{C}{J}\left(\frac{J}{2\pi T}\right)^{1-2\Delta}e^{-i\theta}\frac{\Gamma(\Delta-\frac{i\omega}{2\pi T})}{\Gamma(1-\Delta-\frac{i\omega}{2\pi T})},
\eea
where $C$ is a constant. $\delta G_R(\omega)$ is given by
\bea
\label{eq:deltaGR1}
\delta G_R(\omega)\propto f_R(\omega)G_R^c(\omega),
\eea
which is a correction to the conformal Green's function through perturbation theory. $f_R(\omega)$ is introduced as
\bea
f_R(\omega)=\frac{(2\pi)^{h-2}{\rm cos}\frac{\pi h}{2}\Gamma(h)^2}{{\rm cos}(\pi h)\Gamma(2h-1)}\left[\frac{\Gamma(h)}{\Gamma(1-h)}\left(e^{2i\theta}-\frac{{\rm sin}(\frac{\pi h}{2}-2\pi\Delta)}{{\rm sin}\frac{\pi h}{2}}\right)J_h(\omega)-(h\rightarrow 1-h)\right],\nonumber
\eea
where the function $J_h$ is defined as
\begin{equation}
    J_h=\Gamma(1-\Delta-i\omega)\Gamma(1+h-2\Delta)\Gamma(2\Delta)_3{\bf F}_2\left(
\begin{gathered}
h,h,1+h-2\Delta\\
2h,1+h-\Delta-\frac{i\omega}{2\pi T}
\end{gathered}
; \,1\right),
\end{equation}
where $_3{\bf F}_2$ is the regularized hypergeometric function. For simplicity, we just display the part of Eq.~\eqref{eq:deltaGR1} which  represents the relation between frequency $\omega$ and dimension $h$ in
\bea
\label{eq:deltaGR2}
\delta G_R(\mathfrak{w})\propto A(h)\bigg[\frac{\Gamma(h-1-\Delta-i\mathfrak{w})}{\Gamma(h-\Delta-i\mathfrak{w})}-(h\rightarrow 1-h)\bigg],
\eea
where $A(h)$ is just the coefficient and does not affect these positions. Note that we shift the frequency $\frac{i\omega}{2\pi T}\rightarrow \frac{i\omega}{2\pi T}+2$ and use scaled frequency $\mathfrak{w}:=\frac{\omega}{2\pi T}$ in Eq.~\eqref{eq:deltaGR2}. It is easy for us to obtain the common poles and zeros of the two terms from Eq.~\eqref{eq:deltaGR2}:
\bea
\label{eq:new3}
\left\{
\begin{aligned}
&h-1-\Delta-i\mathfrak{w}=0,-1,-2,-3,\cdots\quad{\rm and}\\
& -h-\Delta-i\mathfrak{w}=0,-1,-2,-3,\cdots \quad\text{(lines of pole)}\,,\\
&h-\Delta-i\mathfrak{w}=0,-1,-2,-3,\cdots\quad\quad{\rm and}\\
& 1-h-\Delta-i\mathfrak{w}=0,-1,-2,-3,\cdots\quad\text{(lines of zero)}\,.
\end{aligned}
\right.
\eea
We consider the large $q$ limit so that $\Delta=1/q$ will become 0 in the SYK model. It is crucial to acknowledge that the Gamma function becomes infinite at negative integers. Consequently, pole-skipping points can be identified at these specific locations, as the corresponding numerator and denominator cannot be mutually canceled out.\\
\indent One can obtain a pole-skipping point in $(\mathfrak{w},h)$ space by finding an intersection between a line of zero and a line of poles ($h$ is a positive value). The following are the resulting pole-skipping points:
\begin{equation}
\label{eq:pspresult2}
\begin{split}
%\mathfrak{w}&=-\frac{i}{2}, \quad h=\frac{1}{2}\,;\\
\mathfrak{w}&=-i, \quad h=1\,;\\
%\mathfrak{w}&=-\frac{3i}{2}, \quad h=\frac{1}{2},\frac{3}{2}\,;\\
\mathfrak{w}&=-2i, \quad  h=1,2\,;\\
%\mathfrak{w}&=-\frac{5i}{2}, \quad h=\frac{1}{2},\frac{3}{2},\frac{5}{2}\,;\\
\mathfrak{w}&=-3i, \quad h=1,2,3\,;\\
&\qquad \qquad \vdots
\end{split}
\end{equation}
\begin{figure}[]
%begin{tabular}{cc}
\begin{minipage}{0.48\linewidth}
\centerline{\includegraphics[width=7.5cm]{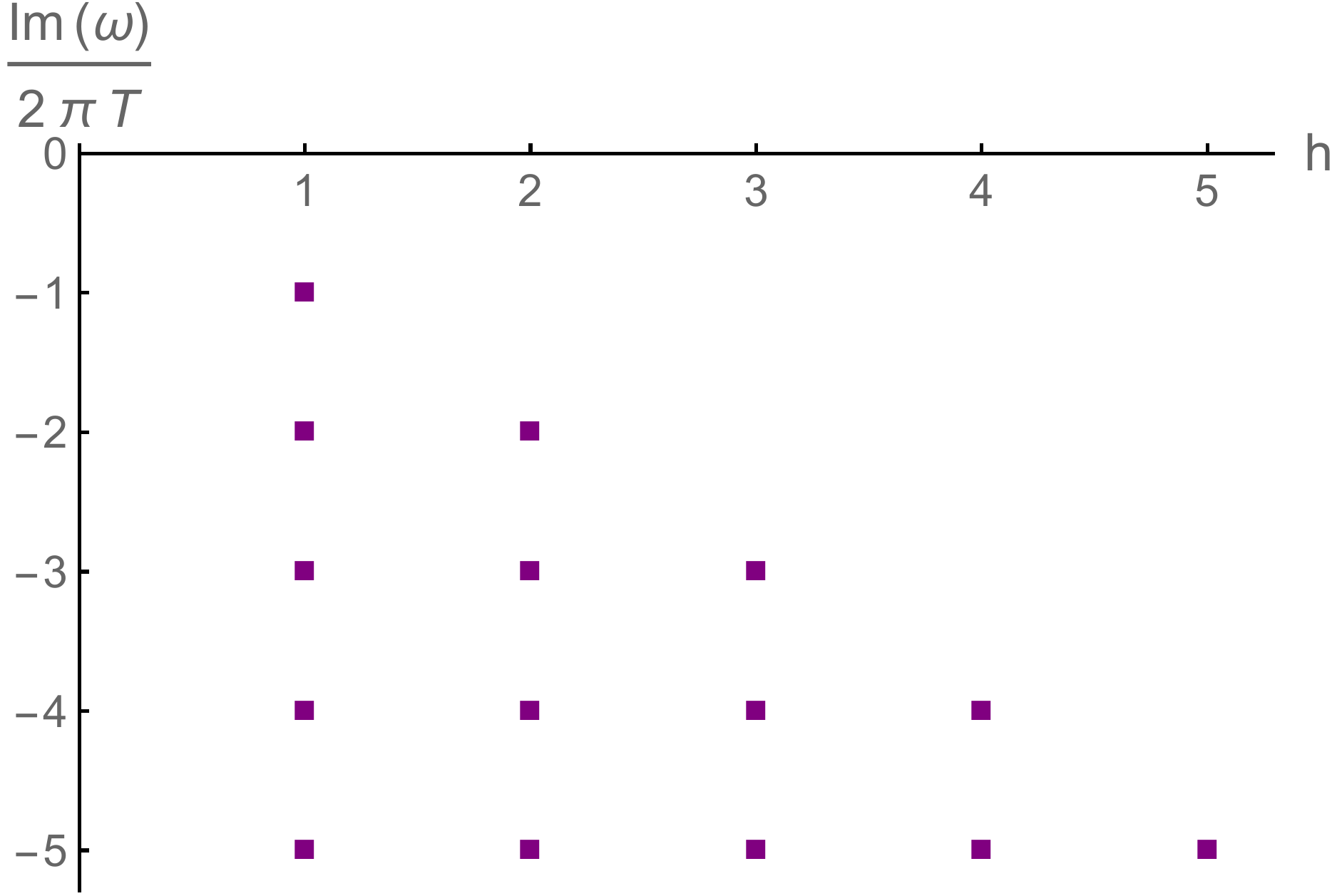}}
\centerline{(a)}
\end{minipage}
\hfill
\begin{minipage}{0.48\linewidth}
\centerline{\includegraphics[width=7.5cm]{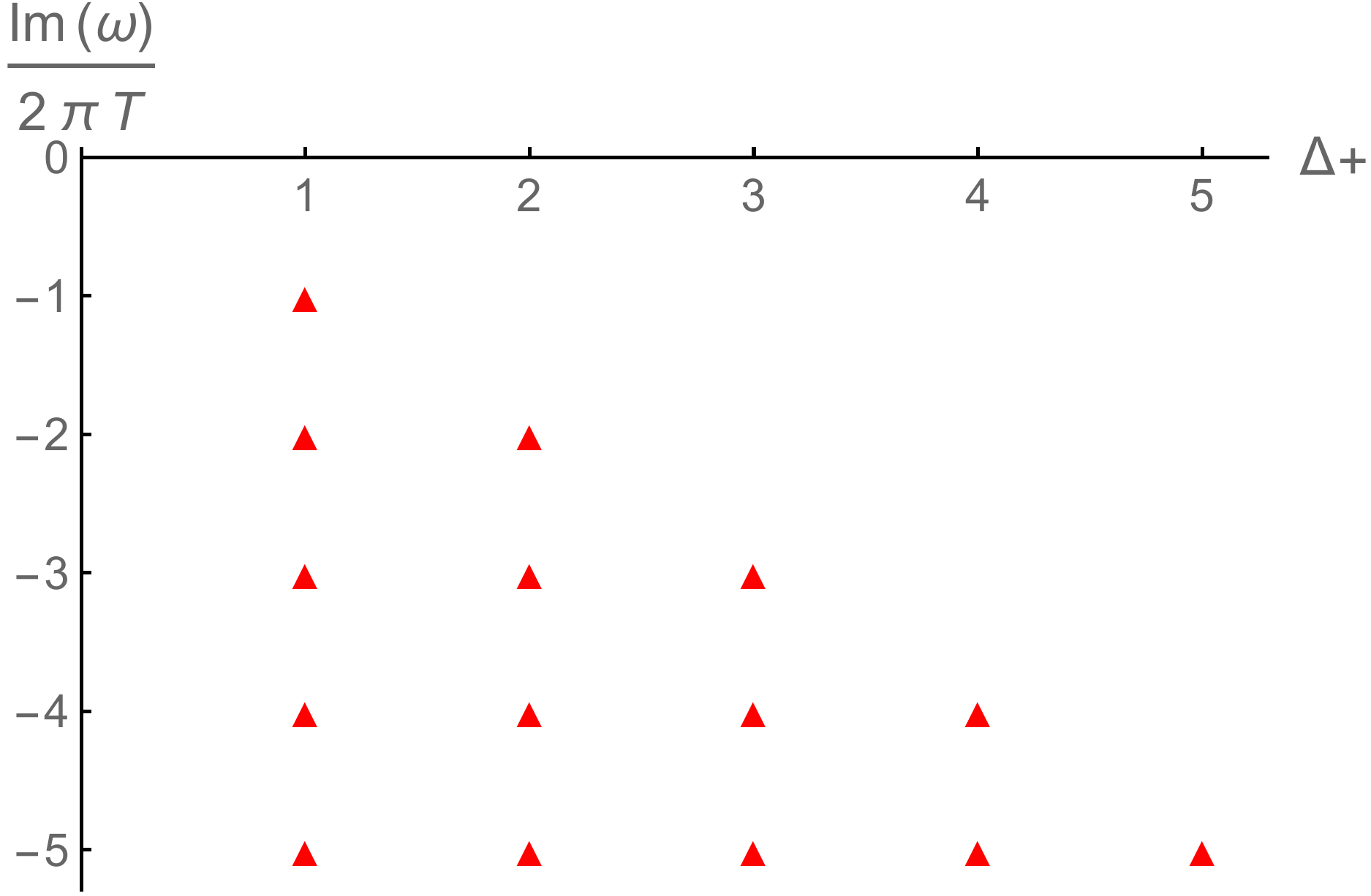}}
\centerline{(b)}
\end{minipage}
\vfill
%end{tabular}
\caption{\label{fig:Figure3} (a) The pole-skipping points of Majorana SYK model in space $(\mathfrak{w},h)$;\; (b) The pole-skipping points of the bulk scalar field in space $(\mathfrak{w},\Delta_+)$.}
\end{figure}
The result \eqref{eq:pspresult2} is the same as the pole-skipping points for standard quantization~\eqref{eq:pspJT} which we obtained in the JT gravity. The figure~\ref{fig:Figure3} shows a graphical comparison between the pole-skipping points given by \eqref{eq:pspresult2} and those obtained in the JT gravity as shown in \eqref{eq:pspJT}.

\begin{figure}[]
%begin{tabular}{cc}
\begin{minipage}{0.48\linewidth}
\centerline{\includegraphics[width=7.5cm]{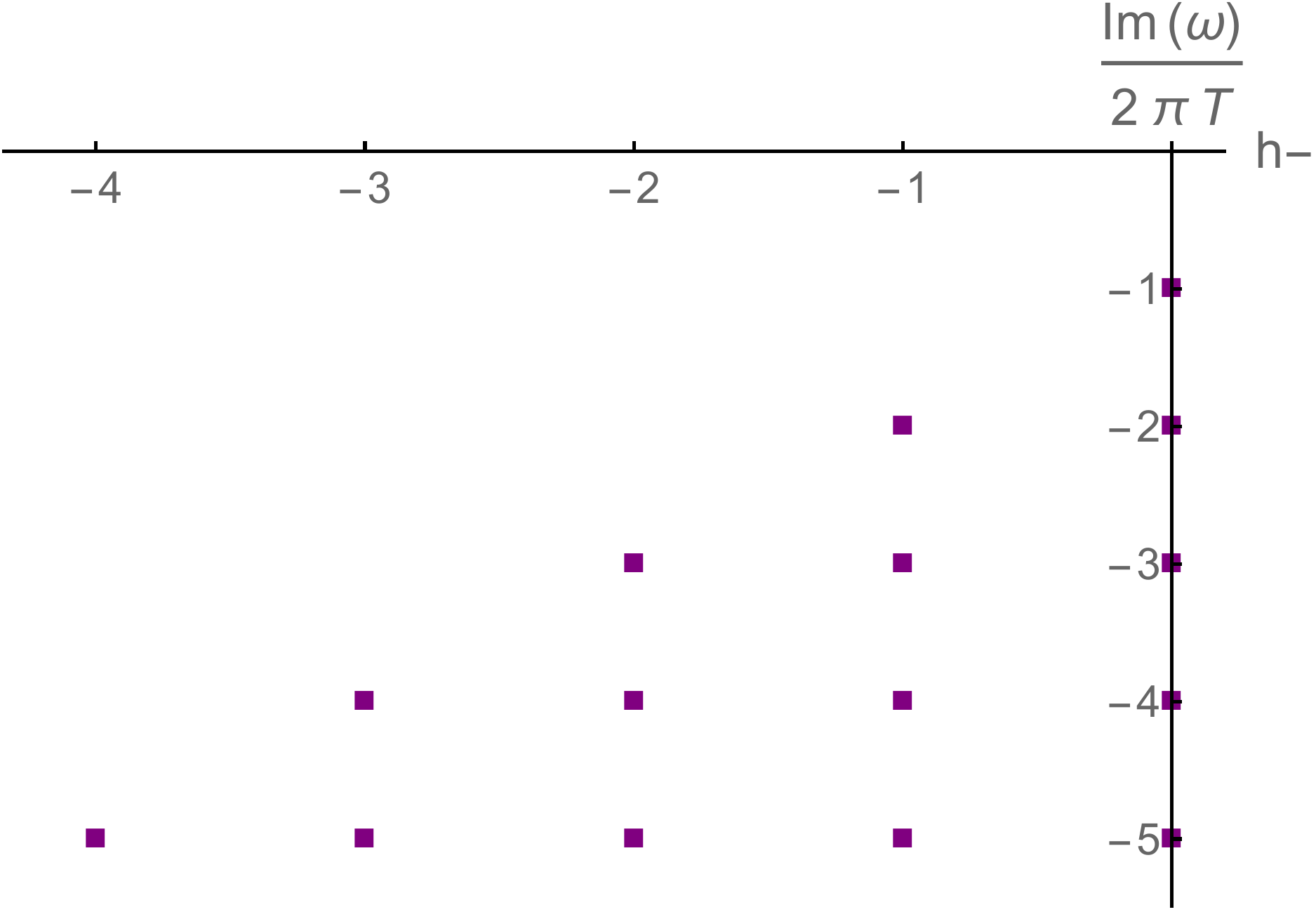}}
\centerline{(a)}
\end{minipage}
\hfill
\begin{minipage}{0.48\linewidth}
\centerline{\includegraphics[width=7.5cm]{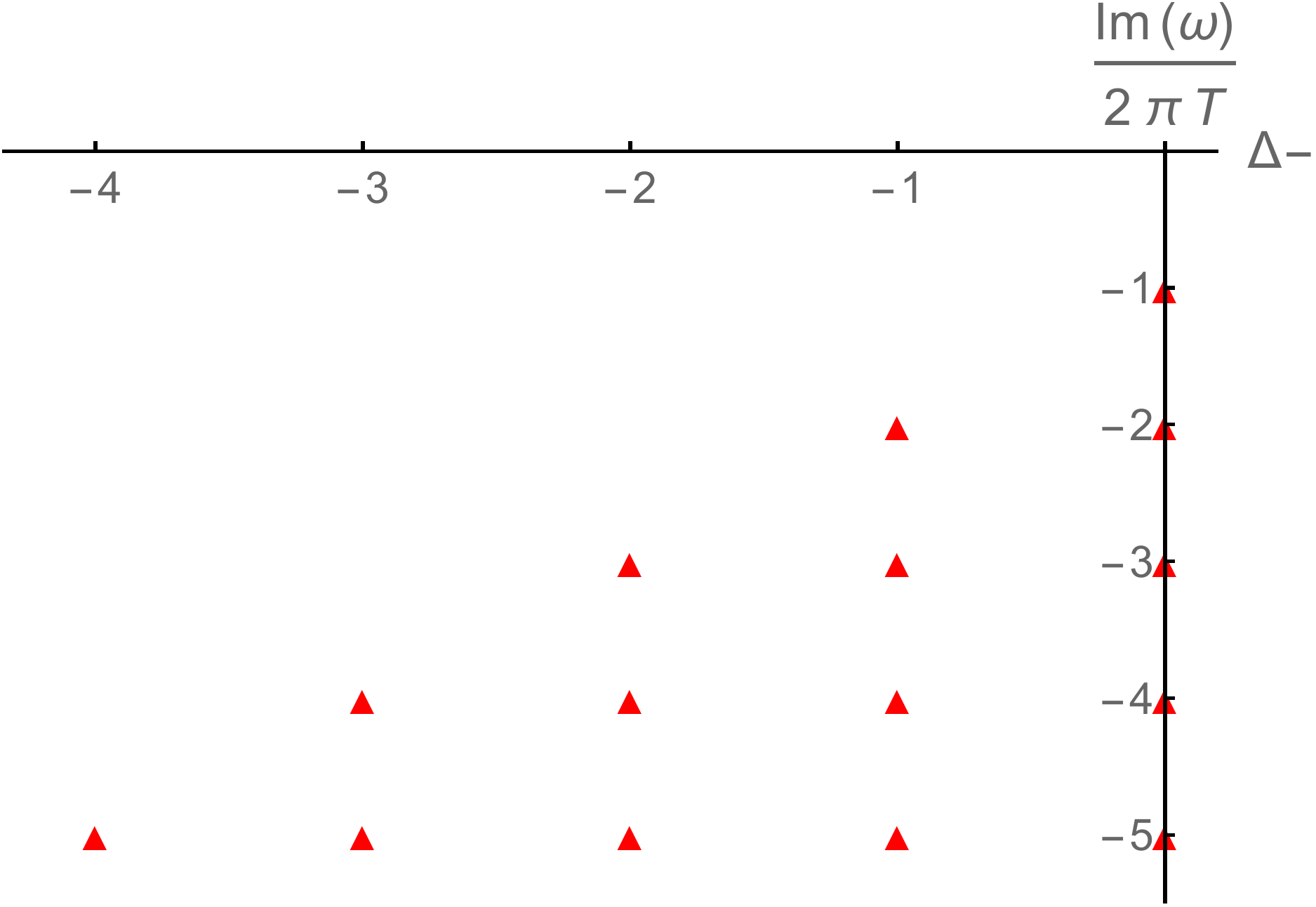}}
\centerline{(b)}
\end{minipage}
\vfill
%end{tabular}
\caption{\label{fig:Figure4} (a) The pole-skipping points of the Majorana SYK model in space $(\mathfrak{w},h_-)$;\; (b) The pole-skipping points of the bulk scalar field with alternative quantization in space $(\mathfrak{w},\Delta_-)$.}
\end{figure}

There are additional pole-skipping points in Eq.~\eqref{eq:deltaGR2}. The expression~\eqref{eq:deltaGR2} is invariant under $h\rightarrow 1-h$ up to sign and the function $A(h)$. Note that the sign and $A(h)$ are not important for pole-skipping. Thus the additional pole-skipping points are obtained after taking replacement $h\rightarrow 1-h$ in Eq.~\eqref{eq:pspresult2}. This transformation $h\rightarrow 1-h$ is similar to transformation $\Delta_+\rightarrow 1-\Delta_+$ for alternative quantization, so we define a new dimension $h_-=1-h$ of operators in SYK corresponding to the boundary scaling dimension $\Delta_-=1-\Delta_+$. The pole-skipping points in plane $(\mathfrak{w},h_-)$ are
\begin{equation}
\label{eq:52}
\begin{split}
\mathfrak{w}&=-i, \quad h_-=0\,;\\
%\mathfrak{w}&=-\frac{3i}{2}, \quad h_-=-\frac{1}{2}\,;\\
\mathfrak{w}&=-2i, \quad  h_-=0,-1\,;\\
%\mathfrak{w}&=-\frac{5i}{2}, \quad h_-=-\frac{1}{2},-\frac{3}{2}\,;\\
\mathfrak{w}&=-3i, \quad h_-=0,-1,-2\,;\\
%\mathfrak{w}&=-\frac{7i}{2}, \quad h_-=-\frac{1}{2},-\frac{3}{2},-\frac{5}{2}\,;\\
&\qquad \qquad \vdots
\end{split}
\end{equation}
The result \eqref{eq:52} is the same as the pole-skipping points for alternative quantization~\eqref{eq:pspJTalt2}. We draw figure~\ref{fig:Figure4} of pole-skipping points in space $(\mathfrak{w},h_-)$ versus pole-skipping points in space $(\mathfrak{w},\Delta_-)$, and show that they are the same. So our assumption that $h_-$ corresponds to $\Delta_-$ is reasonable.
%In bulk side, the relation between two scaling dimensions in the 2-dimensional bulk scalar case also is $\Delta_-=1-\Delta_+$. And the pole-skipping points will not change when we replace $\Delta_+\rightarrow \Delta_-$ in bulk field \cite{Yuan1,Ahn}. Now we define a new variable $h_-=1-h$. In that way $h$ and $h_-$ are defined in the Majorana SYK model just as $\Delta_+$ and $\Delta_-$ are defined as scaling dimensions in the bulk scalar field.\\
%\indent Because $h$ is a positive number with a minimum value of one, $h_-=1-h$ is just a non-positive number. We calculate the pole-skipping points in large $q$ limit when we plug $h_-$ into equation Eq.\eqref{eq:deltaGR2}

%This results are the same as the pole-skipping points we obtained from the retarded Green's function \eqref{eq:deltaGR2} of bulk scalar field when change $\Delta_+\rightarrow\Delta_-$
%\begin{equation}
%\begin{split}
%\label{eq:53}
%&\mathfrak{w}=-i,\quad \Delta_-=0;\\
%&\mathfrak{w}=-2i,\quad \Delta_-=0,-1;\\
%&\mathfrak{w}=-3i,\quad \Delta_-=0,-1,-2;\\
%&\mathfrak{w}=-4i,\quad \Delta_-=0,-1,-2,-3;\\
%&\mathfrak{w}=-5i,\quad \Delta_-=0,-1,-2,-3,-4\,.
%\end{split}
%\end{equation}

\section{Comparison of pole-skipping in charged JT gravity and complex SYK model}
\label{sec:3}
\quad We perceive that this AdS/CFT correspondence is also applicable in the complex SYK model. We will also show this holography through the consistency of pole-skipping of Green's function at charged JT gravity and the complex SYK model.
\subsection{Pole-skipping in charged JT gravity}
\label{sec:cJT}
\quad We will consider the pole-skipping on the boundary of the bulk theory and we will justify the reasons later. We consider the following quadratic action for the complex scalar field~\cite{Thomas}
\bea
S=-\int d^2x\sqrt{-g}\big[(D^\mu\Psi)^\ast D_\mu\Psi+m^2\Psi^\ast\Psi\big]
\eea
with $D_\mu=\partial_\mu-i e A_\mu$. The black hole solution is
\bea
\label{eq:58}
ds^2=\frac{L^2}{u^2}\left(-\left(1-\frac{u^2}{u_0^2}\right)dt^2+\frac{d u^2}{1-\frac{u^2}{u_0^2}}\right)
\eea
with
\bea
\label{eq:61}
A=\epsilon\bigg(\frac{1}{u}-\frac{1}{u_0}\bigg)dt,
\eea
where $u_0$ is the horizon radius, $L$ is the curvature radius of AdS$_2$, and $\epsilon$ is a constant related to chemical potential. Note that we have neglected the backreaction of the gauge field on the background metric. The Hawking temperature is $T=\frac{1}{2\pi u_0}$. Substituting complex scalar field $\Psi(t,u)=e^{-i\omega t}\Psi(\omega,u)$ into the wave equation in the geometry \eqref{eq:58}~\cite{Faulkner}
\bea
\label{eq:66}
\partial^2_u\Psi+\frac{2u}{u^2-u_0^2}\partial_u\Psi+\bigg(-\frac{m^2L^2}{u^2-\frac{u^4}{u_0^2}}+\frac{\big[\omega+\epsilon (\frac{\alpha}{u}-\frac{\alpha}{u_0})\big]^2}{1-\frac{2u^2}{u_0^2}+\frac{u^4}{u_0^4}}\bigg)\Psi=0.
\eea
The parameter $\alpha$ is the coupling strength of the Maxwell field interacting with the scalar field. Considering the asymptotic behavior of $\Psi(t,u)$ as $u\rightarrow 0$, we can calculate the conformal dimension of the scalar field $\Psi(t,u)$ from Eq.~\eqref{eq:66}. This is a homogeneous equation, which can be solved by a power function. Let $\Psi\sim u^{\Delta}$, in the case of the asymptotic boundary ($u\rightarrow 0$), Eq.~\eqref{eq:66} reduced to
\bea
\Delta(1-\Delta)+m^2L^2-\alpha^2\epsilon^2=0
\eea
which implies two results for $\Delta$
\bea
\Delta_\pm=\frac{1}{2}\pm\sqrt{\frac{1}{4}+m^2L^2-\alpha^2\epsilon^2}.
\eea
Introducing the relation
\bea
\Delta_+=\frac{1}{2}+\nu,\quad\nu=\sqrt{\frac{1}{4}+m^2L^2-\alpha^2\epsilon^2},\quad{\rm and}\;\;\Delta_-=\frac{1}{2}-\nu=1-\Delta_+,
\eea
the asymptotic behavior of two linearly independent solutions can be obtained as
\bea
\label{eq:67}
\Psi(u)\sim\big(\frac{1}{u}-\frac{1}{u_0}\big)^{-\frac{1}{2}\mp\nu}\big(\frac{u_0+u}{u_0-u}\big)^{\frac{i\omega u_0}{2}-i\alpha\epsilon}\,_{2}F_1\bigg[\frac{1}{2}\pm\nu+i\omega u_0-i\alpha\epsilon,\frac{1}{2}\pm\nu-i\alpha\epsilon,1\pm 2\nu;\frac{2u}{u-u_0}\bigg].\nonumber\\
\eea
The retarded Green's function of $\Psi$ around the asymptotic boundary can be obtained from Eq.~\eqref{eq:67}
\bea
\label{eq:59}
G_R(\omega)=(4\pi T)^{2\Delta_+-1}\frac{\Gamma(1-2\Delta_+)\Gamma(\Delta_+-\frac{i\omega}{2\pi T}+i\alpha\epsilon)\Gamma(\Delta_+-i\alpha\epsilon)}{\Gamma(2\Delta_+-1)\Gamma(1-\Delta_+-\frac{i\omega}{2\pi T}+i\alpha\epsilon)\Gamma(1-\Delta_+-i\alpha\epsilon)}.
\eea

Here we prove that the retarded Green's function \eqref{eq:17} is the neutral limit of Eq.~\eqref{eq:59}. We use the Legendre duplication formula $\Gamma(z)\Gamma(z+\frac{1}{2})=2^{1-2z}\pi^{1/2}\Gamma(2z)$ to make $\Gamma(\frac{1}{2}-\Delta_+)=4^{\Delta_+}\pi^{1/2}\frac{\Gamma(1-2\Delta_+)}{\Gamma(1-\Delta_+)}$ and $\Gamma(-\frac{1}{2}+\Delta_+)\Gamma(\Delta_+)=2^{2-2\Delta_+}\pi^{1/2}\Gamma(2\Delta_+-1)$. The bulk retarded Green's function \eqref{eq:17} of the scalar field in JT gravity, obtained in subsection~\ref{sec:ScalarBoundary}, becomes
\bea
\label{eq:new5}
G_R&=&(\pi T)^{2\Delta_+-1}\frac{\Gamma(\Delta_{+}-\frac{i\omega}{2\pi T})}{\Gamma(1-\Delta_{+}-\frac{i\omega}{2\pi T})}\frac{\Gamma(\frac{1}{2}-\Delta_{+})}{\Gamma(-\frac{1}{2}+\Delta_{+})}\nonumber\\
&=&(\pi T)^{2\Delta_+-1}4^{\Delta_+}\sqrt{\pi}\frac{\Gamma(\Delta_{+}-\frac{i\omega}{2\pi T})\Gamma(1-2\Delta_{+})\Gamma(\Delta_+)}{\Gamma(1-\Delta_{+}-\frac{i\omega}{2\pi T})\Gamma(1-\Delta_{+})\Gamma(-\frac{1}{2}+\Delta_{+})\Gamma(\Delta_+)}\nonumber\\
&=&(4\pi T)^{2\Delta_+-1}\frac{\Gamma(\Delta_{+}-\frac{i\omega}{2\pi T})\Gamma(1-2\Delta_{+})\Gamma(\Delta_+)}{\Gamma(1-\Delta_{+}-\frac{i\omega}{2\pi T})\Gamma(1-\Delta_{+})\Gamma(2\Delta_+-1)}.
\eea
It is easy to see that the retarded Green's function \eqref{eq:59} will become Eq.~\eqref{eq:new5} when $\alpha=0$.

Let's now redirect our attention to the discussion of Eq.~\eqref{eq:59}. We set $\alpha=1$ hereafter for simplicity, and we regard $i\big(\frac{\omega}{2\pi T}-\epsilon\big)$ as a whole term. The pole-skipping points are located in
\begin{equation}
\begin{split}
\label{eq:pspcJT}
%\left(\frac{\omega}{2\pi T}-\epsilon\right)&=-\frac{i}{2}, \quad \Delta_+=\frac{1}{2};\\
\left(\frac{\omega}{2\pi T}-\epsilon\right)&=-i,\quad \Delta_+=1;\\
%\left(\frac{\omega}{2\pi T}-\epsilon\right)&=-\frac{3i}{2}, \quad \Delta_+=\frac{1}{2},\frac{3}{2};\\
\left(\frac{\omega}{2\pi T}-\epsilon\right)&=-2i,\quad \Delta_+=1,2;\\
%\left(\frac{\omega}{2\pi T}-\epsilon\right)&=-\frac{5i}{2}, \quad \Delta_+=\frac{1}{2},\frac{3}{2},\frac{5}{2};\\
\left(\frac{\omega}{2\pi T}-\epsilon\right)&=-3i,\quad \Delta_+=1,2,3;\\
&\qquad \qquad \vdots
\end{split}
\end{equation}

Note that in this subsection, we have evaluated the pole-skipping points on the boundary of the bulk, unlike in the previous section where we also evaluated them near the event horizon. This is because the scalar potential vanishes at the event horizon, which makes it difficult to observe the effect of charge on the pole-skipping phenomenon. However, this approach can still capture the main physics compared to the pole-skipping points in the complex SYK model. We will see this in the next subsection.
\subsection{Pole-skipping in complex SYK model}
\label{sec:complexSYK}
\quad The complex SYK model describes $N\gg1$ complex fermions with random interactions \cite{Subir,Gu,Maria}. The Hamiltonian is \cite{Subir,Maria}
\bea
H=\frac{1}{2N^{3/2}}\sum^N_{i,j,k,l=1}J_{ij;kl}c^\dagger_ic^\dagger_jc_kc_l-\mu\sum_ic^\dagger_ic_i\,,
\eea
where $c_i$ is complex fermion obeying the anticommutation relations, and the $J_{ij;kl}$ are complex, independent Gaussian random couplings with zero mean. This system also has a conserved $U(1)$ density $\mathcal{Q}$ which is related to the average fermion number as $\mathcal{Q}=\frac{1}{N}\sum_i<c^\dagger_ic_i>$.

The conformal two-point function of the complex SYK model is given as~\cite{Gu}
\bea
G^c(\tau_1,\tau_2)=-b\frac{{\rm sgn}(\tau_{12})}{\vert \tau_{12}\vert^{2\Delta}}e^{2\pi\mathcal{E}(\frac{1}{2}-\frac{\tau}{\beta})},
\eea
where $\mathcal{E}$ is the dimensionless spectral asymmetry parameter which implicitly depends on chemical potential. The conformal effective action of the complex SYK model is
\bea
S_{\rm CFT}=-\frac{N}{2}{\rm log}\;{\rm det}(-\Sigma)-\frac{N}{2}\int d\tau_1d\tau_2\bigg[\Sigma(\tau_1,\tau_2)G(\tau_1,\tau_2)+\frac{1}{q}\big( -G(\tau_1,\tau_2)G(\tau_2,\tau_1)\big)^{\frac{q}{2}}\bigg].\nonumber
\eea
Similar to the Majorana SYK model, the complex SYK model can also be described as a conformal field theory perturbed by an infinite set of irrelevant primary operators. The conformal perturbation theory is still valid for the complex SYK model
\bea
S_{{\rm SYK}}=S_{{\rm CFT}}+\sum_h g_h\int d\tau\; \langle\mathcal{O}^{A/S}_h(\tau)\rangle-\frac{1}{2}\sum_{h,h'} g_h g_{h'}\int d\tau_1d\tau_2\;\langle \mathcal{O}^{A/S}_h(\tau_1)\mathcal{O}^{A/S}_{h'}(\tau_2)\rangle+\dots.\nonumber.
\eea
The difference from Majorana SYK is that there are two kinds of bilinear primary operators, $\mathcal{O}^A_h(\tau)$ and $\mathcal{O}^S_h(\tau)$, representing particle-hole symmetry and asymmetry respectively~\cite{Maria,Gu}. These two irrelevant operators can be schematically represented as $\mathcal{O}^A_{h_n}=f^\dagger_i\partial^{1+2n}_\tau f_i$ and $\mathcal{O}^S_{h_n}=f^\dagger_i\partial^{2n}_\tau f_i$ for nonnegative integer $n$ \cite{Kitaev,Polchinski,Maria,Gross2}. The retarded Green's function in frequency space can be expanded as
\bea
G_R(\omega)=G_R^c(\omega)+\delta G^{A/S}_R(\omega)\,.
\eea
Here $G_R^c(\omega)$ denotes conformal part and its expression is
\bea
G_R^c(\omega)=-i\frac{C}{J}\left(\frac{J}{2\pi T}\right)^{1-2\Delta}e^{-i\theta}\frac{\Gamma(\Delta-\frac{i\omega}{2\pi T}+i\mathcal{E})}{\Gamma(1-\Delta-\frac{i\omega}{2\pi T}+i\mathcal{E})}\,.
\eea
$\delta G^{A/S}_R(\omega)$ can be divided into two part as $\delta G^{A/S}_R(\omega)=\delta G^{A}_R(\omega)+\delta G^{S}_R(\omega)$ correspond to operators $\mathcal{O}^A$ and $\mathcal{O}^S$, where $\delta G^{A/S}_R(\omega)$ is
\begin{equation}
\label{eq:54}
\begin{split}
\delta G^{A/S}_R(\omega)\propto f^{A/S}_R(\omega)G_R^c(\omega)\,,
\end{split}
\end{equation}
and $f^{A/S}_R(\omega)$ is given as~\cite{Maria}
\begin{equation}
\begin{split}
&f^{A}_R(\omega)=\frac{(2\pi)^{h-2}{\rm cos}\frac{\pi h}{2}\Gamma(h)^2}{{\rm cos}(\pi h)\Gamma(2h-1)}\bigg[\frac{\Gamma(h)}{\Gamma(1-h)}\bigg(e^{2i\theta}-\frac{{\rm sin}(\frac{\pi h}{2}-2\pi\Delta)}{{\rm sin}\frac{\pi h}{2}}\bigg)J_h(\omega)-(h\rightarrow 1-h)\bigg]\,,\\
&f^{S}_R(\omega)=\frac{i(2\pi)^{h-2}{\rm sin}\frac{\pi h}{2}\Gamma(h)^2}{{\rm cos}(\pi h)\Gamma(2h-1)}\bigg[\frac{\Gamma(h)}{\Gamma(1-h)}\bigg(\frac{{\rm cos}(\frac{\pi h}{2}-2\pi\Delta)}{{\rm cos}\frac{\pi h}{2}}-e^{2i\theta}\bigg)J_h(\omega)-(h\rightarrow 1-h)\bigg]\,.\nonumber
\end{split}
\end{equation}
With substitution, one can check that $\delta G^{A/S}_R(\omega)$ has the following contribution
\begin{equation}
\label{eq:deltaGcSYK1}
\delta G^{A/S}_R(\omega)\propto (A^A(h)+A^S(h))\bigg[\frac{\Gamma(1+h-\Delta-\frac{i\omega}{2\pi T}+i\mathcal{E})}{\Gamma(2+h-\Delta-\frac{i\omega}{2\pi T}+i\mathcal{E})}-(h\rightarrow 1-h)\bigg]\,,
\end{equation}
where $A^{A/S}(h)$ is a function that depends only on $h$, and it does not affect the locations of pole-skipping. By shifting $i\big(\frac{\omega}{2\pi T}-\mathcal{E}\big)\rightarrow i\big(\frac{\omega}{2\pi T}-\mathcal{E}\big)+2$, the equation \eqref{eq:deltaGcSYK1} becomes
\begin{equation}
\label{eq:deltaGcSYK}
\delta G^{A/S}_R(\omega) \propto (A^A(h)+A^S(h))\bigg[\frac{\Gamma\big(h-1-\Delta-i\big(\frac{\omega}{2\pi T}-\mathcal{E}\big)\big)}{\Gamma\big(h-\Delta-i\big(\frac{\omega}{2\pi T}-\mathcal{E}\big)\big)}-(h\rightarrow 1-h)\bigg]\,.
\end{equation}
We can obtain the common poles and zeros of the two terms from Eq.~\eqref{eq:deltaGcSYK}:
\bea
\label{eq:new4}
\left\{
\begin{aligned}
&h-1-\Delta-i\big(\frac{\omega}{2\pi T}-\mathcal{E}\big)=0,-1,-2,-3,\cdots\quad{\rm and}\\
&-h-\Delta-i\big(\frac{\omega}{2\pi T}-\mathcal{E}\big)=0,-1,-2,-3,\cdots\quad\text{(lines of pole)}\,,\\
&h-\Delta-i\big(\frac{\omega}{2\pi T}-\mathcal{E}\big)=0,-1,-2,-3,\cdots\quad\quad{\rm and}\\
&1-h-\Delta-i\big(\frac{\omega}{2\pi T}-\mathcal{E}\big)=0,-1,-2,-3,\cdots\quad\text{(lines of zero)}\,.
\end{aligned}
\right.
\eea
Our interest is at the large $q$ limit, where $\Delta=1/q$ becomes 0. We can obtain a pole-skipping point in $\big(\frac{\omega}{2\pi T}-\mathcal{E},h\big)$ space by finding an intersection between a line of zeros and a line of poles in Eq.~\eqref{eq:deltaGcSYK}. The result is{
\begin{equation}
\begin{split}
\label{eq:pspcSYK}
%\left(\frac{\omega}{2\pi T}-\mathcal{E}\right)&=-\frac{i}{2}, \quad h=\frac{1}{2};\\
\left(\frac{\omega}{2\pi T}-\mathcal{E}\right)&=-i,\quad h=1;\\
%\left(\frac{\omega}{2\pi T}-\mathcal{E}\right)&=-\frac{3i}{2}, \quad h=\frac{1}{2},\frac{3}{2};\\
\left(\frac{\omega}{2\pi T}-\mathcal{E}\right)&=-2i,\quad h=1,2;\\
%\left(\frac{\omega}{2\pi T}-\mathcal{E}\right)&=-\frac{5i}{2}, \quad h=\frac{1}{2},\frac{3}{2},\frac{5}{2};\\
\left(\frac{\omega}{2\pi T}-\mathcal{E}\right)&=-3i,\quad h=1,2,3;\\
&\qquad \qquad \vdots
\end{split}
\end{equation}
If we interpret $\mathcal{E}$ as $\epsilon$ in~\eqref{eq:pspcJT}, the result Eq.~\eqref{eq:pspcSYK} is the same as the result in charged JT gravity Eq.~\eqref{eq:pspcJT}.

In order to compare $\epsilon$ to $\mathcal{E}$, we need to discuss thermodynamics in charged JT gravity and the complex SYK model. For the thermodynamics of charged black holes in the structure of AdS$_2$ metric
\bea
ds^2=(dr^2-dt^2)/r^2,
\eea
with gauge field $A=(\epsilon/r)dt$. The Bekenstein-Hawking entropy density $\mathcal{S}_{BH}$ at $T=0$ has a relation \cite{Sen1,Sen2}
\bea
\label{eq:zeroTBHentropycJT}
\frac{\partial\mathcal{S}_{BH}}{\partial\mathcal{Q}}=2\pi \epsilon,
\eea
where $\mathcal{Q}$ is a conserved $U(1)$ charge density. As the charge $\mathcal{Q}$  increases, the horizon moves closer to the boundary, and its area $\mathcal{A}_h$ increases. In black hole thermodynamics, the Bekenstein-Hawking entropy density $\mathcal{S}_{BH}$ is related to the area of the horizon via $\mathcal{S}_{BH}=\mathcal{A}_h/(4G_N\mathcal{A}_b)$, where $G_N$ is Newton's constant \cite{Subir}.

In the complex SYK model, the computation of the statistical zero-temperature entropy density $\mathcal{S}$ follows~\cite{Parcollet,Georges} and relies on the thermodynamic Maxwell relation:
\bea
\label{eq:62}
\bigg(\frac{\partial\mathcal{S}}{\partial\mathcal{Q}}\bigg)_T=-\bigg(\frac{\partial\mu}{\partial T}\bigg)_\mathcal{Q}\,.
\eea
At the $T\rightarrow 0$ limit, $\big(\frac{\partial\mu}{\partial T}\big)_\mathcal{Q}=-2\pi \mathcal{E}$. Then the Maxwell relation in Eq.~\eqref{eq:62} leads to
\bea
\label{eq:MaxwellzeroTcSYK}
\bigg(\frac{\partial\mathcal{S}}{\partial\mathcal{Q}}\bigg)=2\pi \mathcal{E}\,.
\eea
The form \eqref{eq:MaxwellzeroTcSYK} in the complex SYK model is the same as Eq.~\eqref{eq:zeroTBHentropycJT} in charged JT gravity. From this comparison, we can interpret the parameter $\epsilon$ as the spectral asymmetry parameter $\mathcal{E}$.\footnote{This interpretation is valid at finite temperature. See~\cite{Subir} for more detail.}

%Ref.~\cite{Subir} also show that the dimensionless parameter $e_d$ which determines the strength of the electric field $A=e_d\big(\frac{1}{r}-\frac{1}{r_h}\big)dt$ is the same as the dimensionless parameter $\mathcal{E}$ appearing as the spectral asymmetry parameter in complex SYK model in the finite-temperature generalization of the AdS$_2$ metric
%\bea
%ds^2=\frac{L^2}{r^2}\left(-\left(1-\frac{r^2}{r_h^2}\right)dt^2+\frac{d r^2}{1-\frac{r^2}{r_h^2}}\right).
%\eea \\
Hence the pole-skipping points \eqref{eq:pspcJT} and \eqref{eq:pspcSYK} show good agreement with each other. Compared to the pole-skipping points in the Majorana SYK model~\eqref{eq:pspresult2}, in the complex SYK model, the spectral asymmetry shifts the pole-skipping frequencies amount of $-\mathcal{E}$.

In the complex SYK model, we can still define a variable $h_-$ corresponding to $\Delta_-$ in charged JT gravity.  Because this relationship is the same as that obtained in the previous section, we do not repeat it here.

\section{Discussion and Conclusion}
\label{sec:Discussion}
\quad In summary, we study the pole-skipping phenomenon on both $(1+1)$-dimensional bulk theories and $(0+1)$-dimensional quantum field theories. We have found a connection between the pole-skipping of gravity and the field theory. In contrast to higher dimensional theories, there is no momentum $k$ degrees of freedom in $(1+1)$-dimensional bulk theories. Instead, we treat mass $m$ as a substitute for momentum for the purpose of identifying the pole-skipping points. Hence our resulting pole-skipping points live in $(\omega,m)$ space instead of $(\omega,k)$ space.

For the JT gravity, we study pole-skipping points by analyzing the near horizon and computing the exact retarded Green's function.
We compare the pole-skipping points in the neutral JT gravity with those in the Majorana SYK model, thanks to the well-known result of the two-point function in the SYK models. The pole-skipping points in space $(\mathfrak{w},h)$ in the Majorana SYK model match those by computing the holographic retarded Green's function in space $(\mathfrak{w},\Delta_+)$ in JT gravity with standard quantization. Moreover, the pole-skipping points in space $(\mathfrak{w},h_-)$ in the Majorana SYK model match those in space $(\mathfrak{w},\Delta_-)$ in JT gravity with alternative quantization. %\textcolor{red}{Interestingly, we find a new dimension $h_-$ of bilinear primary operator. Its physical meaning is consistent with scaling dimension $\Delta_-$ of bulk scalar field in the sense of pole-skipping phenomenon. We have made a correspondence between $\Delta_+$ and $h$, $\Delta_-$ and $h_-$.[YJ: It seems that this discussion is closely related to the alternative quantization in the holographic models. Note that we can also calculate $G^R_{\mathcal{O}\mathcal{O}}$ of a dual operator $\mathcal{O}$ whose scaling dimension $\Delta = \Delta_{-}$. If we want to study this, we also need to figure out the mass regime where the alternative quantization is available.]} \textcolor{blue}{To Yongjun: please rewrite this part.}

Finally, we compute the pole-skipping points in charged JT gravity and in the complex SYK model. The pole-skipping points in these two systems are also in one-to-one correspondence. Compared to the Majorana SYK model, the pole-skipping frequencies are shifted by the spectral asymmetry parameter $\mathcal{E}$, which is analogous to $\epsilon$ with respect to chemical potential in the charged JT gravity. By analyzing zero temperature thermodynamics in charged JT gravity and complex SYK model, we further provide evidence that $\epsilon$ and $\mathcal{E}$ play the same role in each theory.

It would be interesting to study further the pole-skipping of the charged JT gravity and the complex SYK model. The fact that the pole-skipping frequency could be shifted by $\epsilon$ and $\mathcal{E}$ in each model contradicts the well-known result about pole-skipping. As we have just mentioned, the existence of the pole-skipping points in space $(\omega, m)$ instead of space $(\omega, k)$ is a unique property of the phenomenon in two-dimensional gravity. Therefore, we perceive that this type of shift is also a unique property of the pole-skipping phenomenon in charged two-dimensional gravity.

\acknowledgments

We would like to thank Sizheng Cao, Yu-Qi Lei, and Qing-Bing Wang for helpful discussions. This work is partly supported by NSFC, China (No. 12275166 and No.11875184). This work was also supported by the Basic Science Research Program through the National Research Foundation of Korea (NRF) funded by the Ministry of Science, ICT \& Future Planning (NRF- 2021R1A2C1006791), the GIST Research Institute (GRI) and the AI-based GIST Research Scientist Project grant funded by the GIST in 2023.

%\textcolor{red}{\indent We also calculated the pole-skipping points of Dirac field in Appendix~\ref{sec:Dirac}. The locations of the pole-skipping points from near horizon analysis correspond to what we obtained from Green's function on the boundary.[YJ: In my opinion, it is better to do not mention about our result in the Appendix in the discussion section.]}
\appendix
\section{Details of near-horizon expansions}
\label{sec:Details1}
\quad In this appendix, we show the details of the near-horizon expansions of the equations of motion. We can calculate a Taylor series solution to the equation of motion for the scalar mode $\Psi(r)$ when the matrix equation \eqref{eq:4} is satisfied. The first few elements of the matrix are shown below
\bea
&&M_{11}=-\frac{m^2}{2};\nonumber\\
&&M_{21}=0,\quad M_{22}=\frac{f''(r_h)-m^2}{4};\nonumber\\
&&M_{31}=0,\quad M_{32}=\frac{f^{(3)}(r_h)}{12},\quad M_{33}=\frac{m^2-3f''(r_h)}{6r_h};\\
&&M_{41}=0,\quad M_{42}=\frac{f^{(4)}(r_h)}{48},\quad M_{43}=\frac{f^{(3)}(r_h)}{6},\quad M_{43}=\frac{6f''(r_h)-m^2}{8};\nonumber\\
&&M_{51}=0,\quad M_{52}=\frac{f^{(5)}(r_h)}{240},\quad M_{53}=\frac{f^{(4)}(r_h)}{24},\quad M_{54}=\frac{f^{(3)}(r_h)}{4},\quad M_{54}=\frac{10f''(r_h)-m^2}{10}.\nonumber
\eea
\section{Pole-skipping of Dirac field in JT gravity}
\label{sec:Dirac}
\quad As a way to confirm the validity of our holographic computation method, we consider the Dirac field (spin $s=\frac{1}{2}$) in the same background \eqref{eq:2}. The Dirac equation is given as
\bea
\label{eq:19}
(\Gamma^MD_M-m)\psi_{\pm}=0.
\eea
The capital letter $M$ denotes the indices of bulk spacetime coordinates and small letters $a,b$ denote tangent space indices. The covariant derivative of bulk spacetime acting on fermions is defined by $D_M=\partial_M+\frac{1}{4}(\omega_{ab})_M\Gamma^{ab}$, where $\Gamma_{ab}\equiv\frac{1}{2}[\Gamma_a,\Gamma_b]$. $\Gamma_a$ are Gamma matrices which satisfy Grassman algebra $\{\Gamma^a,\Gamma^b\}=2\eta^{ab}$ \cite{N1,Cai}. The spinors are two dimensional $\psi_{\pm}(r,t)=e^{-i\omega t}\left(
  \begin{array}{c}
    \psi_{+}(r)\\
    \psi_{-}(r)\\
  \end{array}
\right)$ and we  use the following gamma matrices representation \cite{Wilczek}
\bea
\label{eq:36}
\Gamma^{\underline{v}}=i\sigma^2,\quad \Gamma^{\underline{r}}=\sigma^3.
\eea
We choose the orthonormal frame to be
\bea
\label{eq:20}
E^{\underline{v}}=\frac{1+f(r)}{2}dv-dr,\quad E^{\underline{r}}=\frac{1-f(r)}{2}dv+dr,
\eea
for which
\bea
\label{eq:21}
ds^2=\eta_{ab}E^aE^b,\quad \eta_{ab}={\rm diag}(-1,1).
\eea
The spin connections for this frame are given by
\bea
\label{eq:37}
\omega_{\underline{rr}}=0,\quad \omega_{\underline{vr}}=-\frac{f'(r)}{2}.
\eea
Using these spin connections \eqref{eq:37} and the Eddington-Finkelstein coordinate \eqref{eq:2} of Schwarzschild-AdS$_2$
\bea
\label{eq:38}
ds^2=-f(r)dv^2+2dvdr,
\eea
one can calculate the Dirac equation to be
\bea
\label{eq:22}
\left\{
\begin{aligned}
&\big(f'(r)-4(m+i\omega)\big)\psi_+(r)+\big(f'(r)-4i\omega\big)\psi_-(r)+2\big(f(r)-1\big)\psi'_-(r)\\
&+2\big(f(r)+1\big)\psi'_+(r)=0,\\
&\big(4i\omega-f'(r)\big)\psi_+(r)-\big(4m+f'(r)-4i\omega\big)\psi_-(r)-2\big(f(r)+1\big)\psi'_-(r)\\
&+2\big(f(r)-1\big)\psi'_+(r)=0.
\end{aligned}
\right.
\eea
\subsection{Pole-skipping: near-horizon analysis}
\label{sec:DiracNear}
\quad
We combine the two equations of \eqref{eq:22} and expand them near the horizon $r_h$. The first-order equation near the horizon is
\bea
\label{eq:23}
1st:\quad(2\pi T-2i\omega-m)\psi_++(2\pi T-2i\omega+m)\psi_-=0.
\eea
We take the value of coefficients $(2\pi T-2i\omega-m)$ and $(2\pi T-2i\omega+m)$ to be 0 and thus there are two independent free parameters $\psi_+$ and $\psi_-$ to this equation. The first-order pole-skipping point is obtained as
\bea
\label{eq:24}
\mathfrak{w}=-\frac{i}{2}, \quad \mathfrak{m}=0, \quad \Delta=\frac{1}{2}.
\eea
We expand the Dirac equation in higher order
\bea
2nd:&&\bigg(\frac{m(\pi T-m)-\frac{m^2}{4\pi T}}{m+2\pi T-2i\omega}\bigg)\psi^0_{+}+\frac{1}{2}\bigg(3-\frac{i\omega}{\pi T}\bigg)\psi^1_{+}=0,\\
3rd:&&\frac{m(4\pi^2T^2-m^2)(m^2+10m\pi T-16\pi^2T^2-2im\omega)}{8\pi T(3\pi T-i\omega)(m+2\pi T-2i\omega)}\psi^0_{+}+\big(5-\frac{i\omega}{\pi T}\big)\psi^2_{+}=0,\\
4th:&&\frac{m(20m^2\pi^2T^2-64\pi^4T^4-m^4)(m^2+14m\pi T-36\pi^2T^2-2im\omega)}{32\pi T(3\pi T-i\omega)(5\pi T-i\omega)(m+2\pi T-2i\omega)}\psi^0_{+}\nonumber\\
&&+\frac{3}{2}\big(7-\frac{i\omega}{\pi T}\big)\psi^3_{+}=0,\\
5th:&&\frac{m(56m^4\pi^2T^2+2304\pi^6T^6-784m^2\pi^4T^4-m^6)(m^2+18m\pi T-64\pi^2T^2-2im\omega)}{192\pi T(3\pi T-i\omega)(5\pi T-i\omega)(7\pi T-i\omega)(m+2\pi T-2i\omega)}\psi^0_{+}\nonumber\\
&&+2\big(9-\frac{i\omega}{\pi T}\big)\psi^4_{+}=0.
\eea
In general bulk dimensions, the mass $m$ of the fermionic field and the scaling dimension $\Delta$ of the dual operator are related via
\bea
\Delta=\frac{d+1}{2}+m.
\eea
In the case of AdS$_2$, $d=1$, and we get the relation $\Delta=\frac{1}{2}+m$. The higher-order pole-skipping points are
\begin{equation}
\label{eq:25}
\begin{split}
\mathfrak{w}&=-\frac{3i}{2}, \quad \mathfrak{m}=-1,0,1, \quad \Delta=-\frac{1}{2},\frac{1}{2},\frac{3}{2}\,;\\
\mathfrak{w}&=-\frac{5i}{2}, \quad \mathfrak{m}=-2,-1,0,1,2, \quad \Delta=-\frac{3}{2},-\frac{1}{2},\frac{1}{2},\frac{3}{2},\frac{5}{2}\,;\\
\mathfrak{w}&=-\frac{7i}{2}, \quad \mathfrak{m}=-3,-2,-1,0,1,2,3, \quad \Delta=-\frac{5}{2},-\frac{3}{2},-\frac{1}{2},\frac{1}{2},\frac{3}{2},\frac{5}{2},\frac{7}{2}\,;\\
&\qquad \qquad \vdots
\end{split}
\end{equation}
We calculate the first few order pole-skipping points and plot them in figure~\ref{fig:Figure2}.
\subsection{Pole-skipping from Green's function}
\label{sec:DiracBoundary}
\quad In this subsection, we also compute the pole-skipping points from the exact retarded Green's function on the boundary. We substitute the metric \eqref{eq:14} into the Dirac equation \eqref{eq:19}
\bea
\label{eq:31}
\left\{
\begin{aligned}
-2iL\omega{\rm Csch}(u)\psi_-(u)+({\rm Coth}(u)-2Lm)\psi_+(u)+2\psi'_+(u)=0,\\
(2Lm+{\rm Coth}(u))\psi_-(u)-2iL\omega{\rm Csch}(u)\psi_+(u)+2\psi'_-(u)=0.
\end{aligned}
\right.
\eea
Combining the two equations of \eqref{eq:31}, we obtain
\bea
\label{eq:26}
\left\{
\begin{aligned}
&4\psi''_+(u)+8{\rm Coth}(u)\psi'_+(u)\\
&+\big(3{\rm Coth}^2(u)-4mL{\rm Coth}(u)+2(\omega^2L^2-1){\rm Csch}^2(u)-4m^2L^2\big)\psi_+(u)=0,\\
&4\psi''_-(u)+8{\rm Coth}(u)\psi'_-(u)\\
&+\big(3{\rm Coth}^2(u)+4mL{\rm Coth}(u)+2(\omega^2L^2-1){\rm Csch}^2(u)-4m^2L^2\big)\psi_-(u)=0.
\end{aligned}
\right.
\eea
We do another change of variable $z={\rm tanh}^2(u)$, and solve these two second order differential equations \eqref{eq:26}. Near the asymptotic boundary ($z\rightarrow 1$), the two spinor components behave as
\bea
\label{eq:27}
\psi_+&\sim&C_1\,_{2}F_1\bigg[-iL\omega,\frac{1}{2}+Lm-i\omega,1-2iL\omega;\frac{2z}{z-1}\bigg]\nonumber\\
&&+(-2)^{2iL\omega}(\frac{z}{z-1})^{2iL\omega}C_2\,_{2}F_1\bigg[iL\omega,\frac{1}{2}+Lm+i\omega,1+2iL\omega;\frac{2z}{z-1}\bigg]\nonumber\\
&\sim&(1-z)^{\frac{1}{4}-\frac{mL}{2}}A+(1-z)^{\frac{3}{4}+\frac{mL}{2}}B\nonumber\\
&=&(1-z)^{\frac{1}{2}-\frac{\Delta}{2}}A+(1-z)^{\frac{1}{2}+\frac{\Delta}{2}}B;
\eea
\bea
\label{eq:28}
\psi_-&\sim&C_1\,_{2}F_1\bigg[-iL\omega,\frac{1}{2}-Lm-i\omega,1-2iL\omega;\frac{2z}{z-1}\bigg]\nonumber\\
&&+(-2)^{2iL\omega}(\frac{z}{z-1})^{2iL\omega}C_2\,_{2}F_1\bigg[\frac{1}{2}-Lm+i\omega,iL\omega,1+2iL\omega;\frac{2z}{z-1}\bigg]\nonumber\\
&\sim&(1-z)^{\frac{1}{4}+\frac{mL}{2}}C+(1-z)^{\frac{3}{4}-\frac{mL}{2}}D\nonumber\\
&=&(1-z)^{\frac{\Delta}{2}}C+(1-z)^{1-\frac{\Delta}{2}}D.
\eea
We choose $A$ as the source and the expectation value as $C$. The retarded Green's function in this case is given by their ratio
\bea
\label{eq:29}
&&G_R\propto i\frac{C}{A}=i\frac{\Gamma(\frac{1}{2}+Lm-i\omega L)\Gamma(\frac{1}{2}- Lm)}{\Gamma(\frac{1}{2}-Lm-i\omega L)\Gamma(\frac{1}{2}+Lm)}=
i\frac{\Gamma(\Delta-i\omega L)}{\Gamma(1-\Delta-i\omega L)}\frac{\Gamma(1-\Delta)}{\Gamma(\Delta)}.
\eea
The poles and zeros of the Green's function are
\bea
\label{eq:35}
\left\{
\begin{aligned}
\frac{1}{2}+Lm-i\omega L=0,-1,-2,-3,\cdots\,, \qquad\text{(lines of pole)}\,,\\
\frac{1}{2}-Lm-i\omega L=0,-1,-2,-3,\cdots\,,\qquad\text{(lines of zero)}\,.
\end{aligned}
\right.
\eea
According to these two equations, we obtain the pole-skipping points
\begin{equation}
\label{eq:30}
\begin{split}
\mathfrak{w}&=-\frac{i}{2}, \quad \Delta=\frac{1}{2}\,;\\
%\mathfrak{w}&=-i, \quad \Delta=0,1\,;\\
\mathfrak{w}&=-\frac{3i}{2}, \quad \Delta=-\frac{1}{2},\frac{1}{2},\frac{3}{2}\,;\\
%\mathfrak{w}&=-2i, \quad  \Delta=-1,0,1,2\,;\\
\mathfrak{w}&=-\frac{5i}{2}, \quad \Delta=-\frac{3}{2},-\frac{1}{2},\frac{1}{2},\frac{3}{2},\frac{5}{2}\,;\\
%\mathfrak{w}&=-3i, \quad \Delta=-2,-1,0,1,2,3\,;\\
%\mathfrak{w}&=-\frac{7i}{2}, \quad \Delta=\Delta=-\frac{5}{2},-\frac{3}{2},-\frac{1}{2},\frac{1}{2},\frac{3}{2},\frac{5}{2},\frac{7}{2}\,;\\
%\mathfrak{w}&=-4i, \quad \Delta=-3,-2,-1,0,1,2,3,4\,;\\
&\qquad \qquad \vdots
\end{split}
\end{equation}
From result \eqref{eq:30}, we can see that the pole-skipping points from the exact retarded Green's function on the boundary are the same as these special points in \eqref{eq:24} and \eqref{eq:25} near the horizon. The comparison of the pole-skipping points obtained by these two methods is shown in figure~\ref{fig:Figure2}.
\begin{figure}[!t]
%begin{tabular}{cc}
\begin{minipage}{0.48\linewidth}
\centerline{\includegraphics[width=7.5cm]{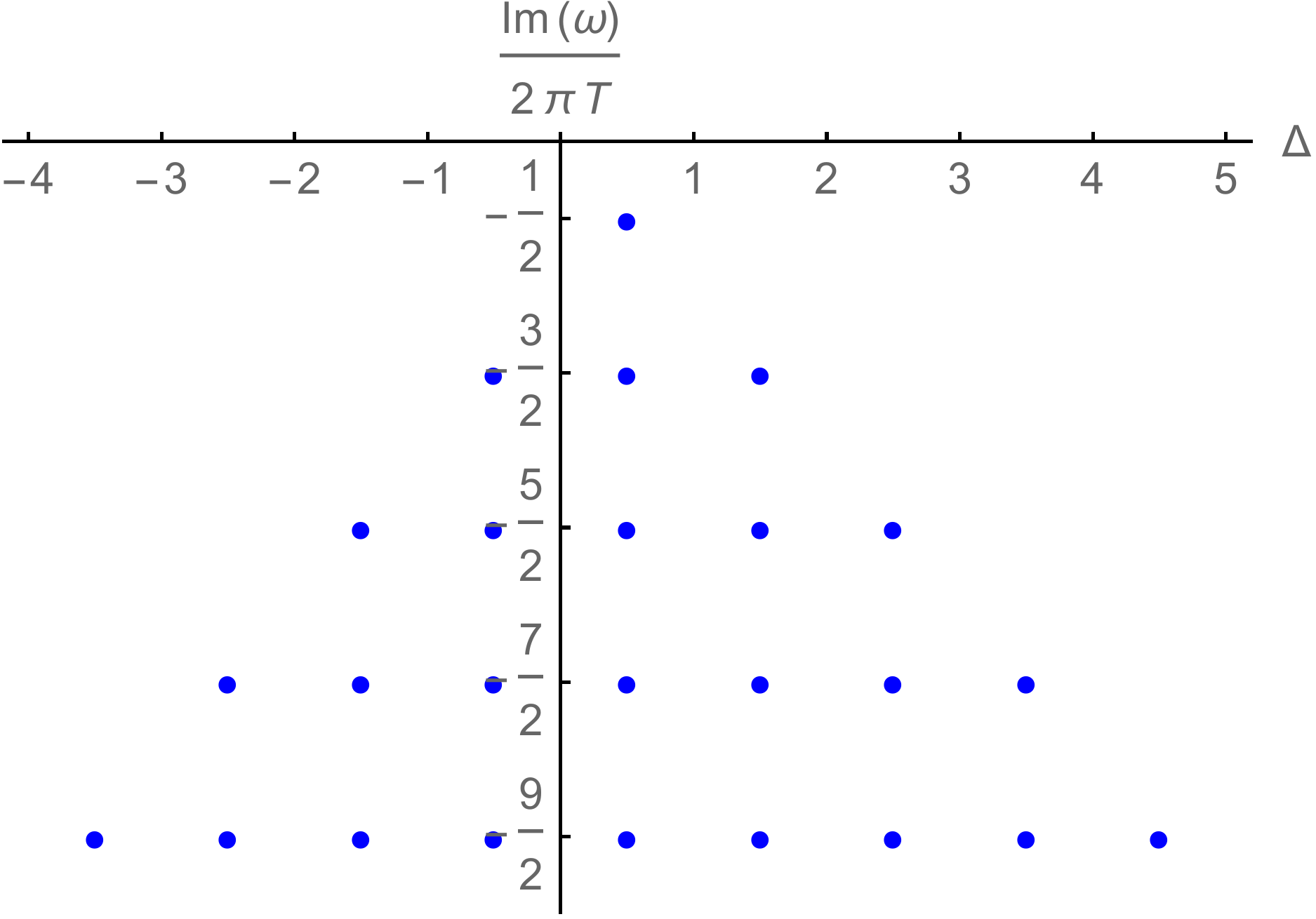}}
\centerline{(a)}
\end{minipage}
\hfill
\begin{minipage}{0.48\linewidth}
\centerline{\includegraphics[width=7.5cm]{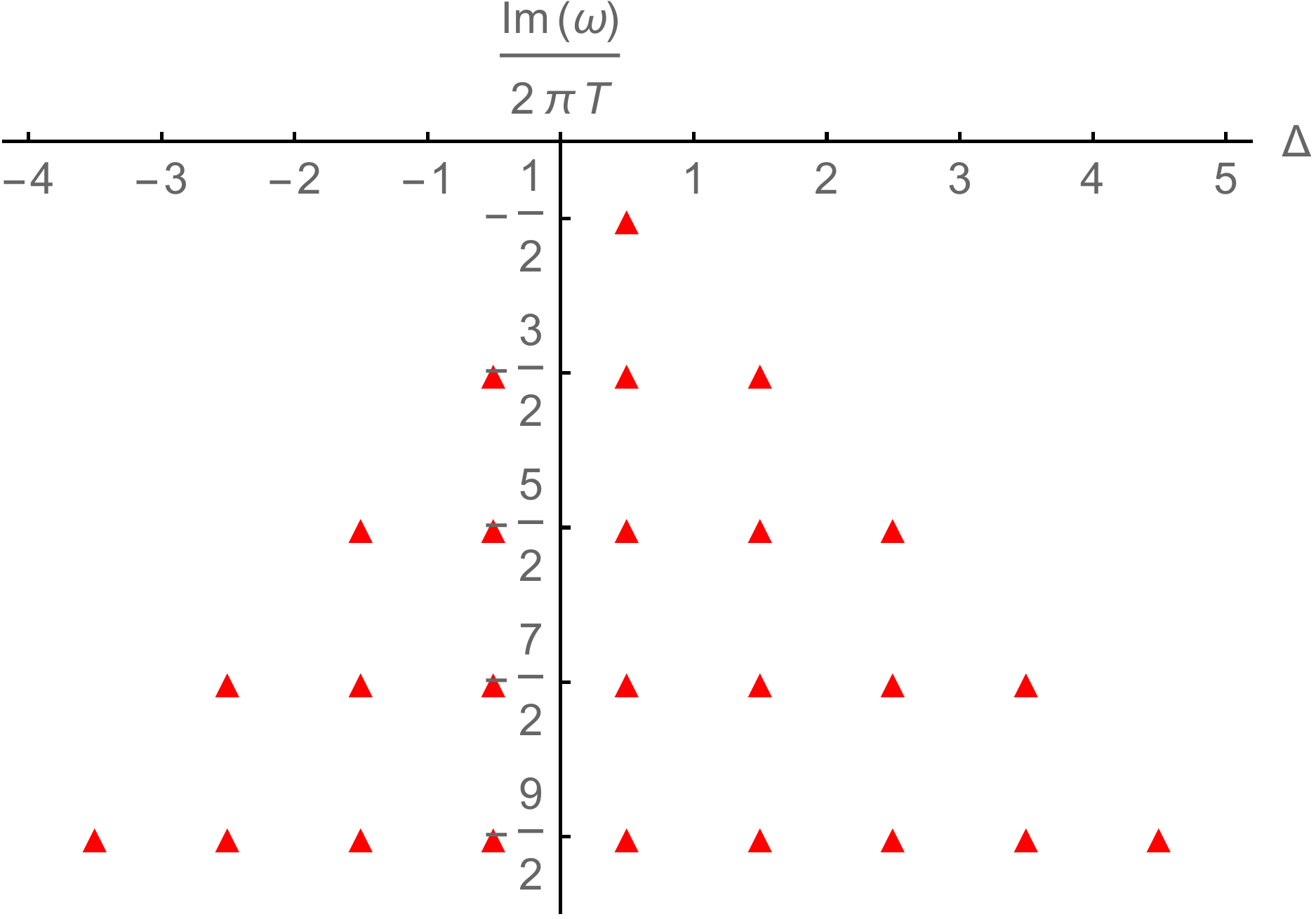}}
\centerline{(b)}
\end{minipage}
\vfill
%end{tabular}
\caption{\label{fig:Figure2} The pole-skipping points \eqref{eq:25} and \eqref{eq:30} of Dirac field in JT black hole background.\; (a) From near the horizon ;\; (b) From boundary retarded Green's function.}
\end{figure}

% The bibliography will probably be heavily edited during typesetting.
% We'll parse it and, using the arxiv number or the journal data, will
% query inspire, trying to verify the data (this will probalby spot
% eventual typos) and retrive the document DOI and eventual errata.
% We however suggest to always provide author, title and journal data:
% in short all the informations that clearly identify a document.


\begin{thebibliography}{99}

\bibitem{maldacena}
J. Maldacena, \emph{The Large-N Limit of Superconformal Field Theories and Supergravity}, \emph{Int. J. Theor. Phys.} {\bf 38} (1999) 1113, arXiv:9711200.
\bibitem{witten1}E. Witten, \emph{Anti-de Sitter space and holography}, \emph{Adv. Theor. Math. Phys.} {\bf 2} (1988) 253, arXiv:9802150.
\bibitem{witten2}E. Witten, \emph{Anti-de Sitter space, thermal phase transition, and confinement in gauge theories}, \emph{Adv. Theor. Math. Phys.} {\bf 2} (1998)  505, arXiv:9803131.
\bibitem{gubser}S. S. Gubser, I. R. Klebanov and A. M. Polyakov, \emph{Gauge theory correlators from noncritical string theory}, \emph{Phys. Lett. B.} {\bf428} (1998) 105, arXiv:9802109.
\bibitem{casalderrey}J. Casalderrey-Solana, H. Liu, D. Mateos, K. Rajagopal and U. A. Wiedemann, \emph{Gauge/String Duality, Hot QCD and Heavy Ion Collisions}, Cambridge Univ. Press (2014), arXiv:1101.0618.
\bibitem{natsuume}M. Natsuume, \emph{AdS/CFT Duality User Guide}, Springer Japan, Tokyo (2015), arXiv:1409.3575.
\bibitem{ammon}M. Ammon and J. Erdmenger, \emph{Gauge/gravity duality: Foundations and applications}, Cambridge Univ. Press (2015).
\bibitem{zaanen}J. Zaanen, Y. W. Sun, Y. Liu and K. Schalm, \emph{Holographic Duality in Condensed Matter Physics}, Cambridge Univ. Press (2015).
\bibitem{hartnoll}S. A. Hartnoll, A. Lucas and S. Sachdev, \emph{Holographic quantum matter}, The MIT Press (2018).
\bibitem{Grozdanov1}S. Grozdanov, K. Schalm, and V. Scopelliti, \emph{Black Hole Scrambling from Hydrodynamics}, Phys. Rev. Lett. {\bf120}, 231601 (2018), arXiv:1710.00921.
\bibitem{Blake1}M. Blake, R. A. Davions, S. Grozdanov, and H. Liu, \emph{Many-body chaos and energy dynamics in holography}, JHEP {\bf2018}, 35 (2018), arXiv:1809.01169.
\bibitem{Grozdanov2}S. Grozdanov, \emph{On the connection between hydrodynamics and quantum chaos in holographic theories with stringy corrections}, JHEP {\bf2019}, 48 (2019), arXiv:1811.09641.
\bibitem{Makoto1}M. Natsuume and T. Okamura, \emph{Holographic chaos, pole-skipping, and regularity}, \emph{Progress of Theoretical and Experimental Physics}, {\bf 1} (2020) 013B07, arXiv:1905.12014.
\bibitem{Makoto2}M. Natsuume and T. Okamura, \emph{Nonuniqueness of Green's functions at special points}, arXiv:1905.12015.
\bibitem{BlakeDavison}M. Blake, R. A. Davison, and D. Vegh, \emph{Horizon constraints on holographic Green's functions}, JHEP {\bf2020}, 77 (2020), arXiv:1904.12883.
\bibitem{Das}S. Das, B. Ezhuthachan and A. Kundu, \emph{Real time dynamics from low point correlators in 2d BCFT}, JHEP {\bf 2019} 141 (2019), arXiv:1907.08763.
\bibitem{Abbasi2} N. Abbasi and S. Tahery, \emph{Complexified quasinormal modes and the pole-skipping in a holographic system at finite chemical potential}, JHEP {\bf 2020} 76 (2020), arXiv:2007.10024.
\bibitem{Abbasi1}
N. Abbasi and J. Tabatabaei, \emph{Quantum chaos, pole-skipping and hydrodynamics in a holographic system with chiral anomaly}, JHEP {\bf2020}, 50 (2020), arXiv:1910.13696.
\bibitem{Choi}C. Choi, M. Mezei and G. S{\'a}rosi, \emph{Pole skipping away from maximal chaos}, JHEP {\bf2021}, 207 (2021), arXiv: 2010.08558.
\bibitem{Karunava}K. Sil, \emph{Pole skipping and chaos in anisotropic plasma: a holographic study}, JHEP {\bf2021}, 232 (2021), arXiv:2012.07710.
\bibitem{Yongjun2}Y.~Ahn, V.~Jahnke, H.~S.~Jeong, K.~Y.~Kim,  K.~S.~Lee, and M.~Nishida, \emph{Classifying pole-skipping points}, arXiv:2010.16166.
\bibitem{Mahdi} M. Atashi and K. Bitaghsir Fadafan, \emph{Holographic pole-skipping of flavor branes}, Journal of Holography Applications in Physics, {\bf2}(2), pp. 39-46 (2022). doi: 10.22128/jhap.2022.519.1020.
\bibitem{Makoto3}M. Natsuume and T. Okamura, \emph{Pole-skipping with finite-coupling corrections}, Phys. Rev. D {\bf100}, 126012 (2019), arXiv:1909.09168.
\bibitem{N1}
N. {\'C}eplak, K. Ramdial, and D. Vegh, \emph{Fermionic pole-skipping in holography}, JHEP {\bf2020}, 203 (2020), arXiv:1910.02975.
\bibitem{Yuan1}
H. Yuan, X. H. Ge, \emph{Pole-skipping and hydrodynamic analysis in Lifshitz, AdS2 and Rindler geometries}, JHEP {\bf2021}, 165 (2021), arXiv:2012.15396.
\bibitem{N2}
N. {\'C}eplak and D. Vegh, \emph{Pole skipping and Rarita-Schwinger fields}, Phys. Rev. D {\bf103}, 106009 (2021), arXiv:2101.01490.
\bibitem{Yuan2}
H. Yuan, X. H. Ge, \emph{Analogue of the pole-skipping phenomenon in acoustic black holes}, Eur. Phys. J. C {\bf82}, 167 (2022), arXiv:2110.08074.
\bibitem{Diandian} D. Wang and Z. Y. Wang, \emph{Pole Skipping in Holographic Theories with Bosonic Fields}, Phys. Rev. Lett. {\bf129}, 231603 (2022), arXiv:2208.01047.
\bibitem{Blake3}
M. Blake, H. Lee, and H. Liu, \emph{A quantum hydrodynamical description for scrambling and many-body chaos}, JHEP {\bf2018} 127 (2018), arXiv:1801.00010.
\bibitem{Jeong}
H. S. Jeong, K. Y. Kim, and Y. W. Sun, \emph{Bound of diffusion constants from pole-skipping points: spontaneous symmetry breaking and magnetic field}, JHEP {\bf 2021}, 105 (2021), arXiv:2104.13084.
\bibitem{Jackiw}
R. Jackiw, \emph{Lower dimensional gravity}, Nucl. Phys. B {\bf252}, 343 (1985).
\bibitem{Teitelboim}
C. Teitelboim, \emph{Gravitation and Hamiltonian structure in two space-time dimensions}, Phys. Lett. B {\bf126}, 41 (1983).
\bibitem{Sachdev}
S. Sachdev and J. Ye, \emph{Gapless spin-fluid ground state in a random quantum Heisenberg magnet}, Physical Review Letters 70 3339 (1993), arXiv:cond-mat/9212030.
\bibitem{Kitaev}
A. Kitaev, \emph{A simple model of quantum holography},\\
http://online.kitp.ucsb.edu/online/entangled15/kitaev/,\\
http://online.kitp.ucsb.edu/online/entangled15/kitaev2/. Talks at KITP, April 7, 2015 and May 27, 2015.
\bibitem{Jensen}
K. Jensen, \emph{Chaos in AdS2 holography}, Phys. Rev. Lett. {\bf117} 111601 (2016), arXiv:1605.06098.
\bibitem{Maldacena:2015waa}
J.~Maldacena, S.~H.~Shenker and D.~Stanford,
\emph{A bound on chaos},
JHEP \textbf{08}, 106 (2016), arXiv:1503.01409.
\bibitem{Maldacena}
J. Maldacena, D. Stanford and Z. Yang, \emph{Conformal symmetry and its breaking in two dimensional Nearly Anti-de-Sitter space}, PTEP {\bf2016}, no. 12, 12C104 (2016), arXiv:1606.01857.
\bibitem{Sarosi}
G. {\'S}arosi, AdS2 holography and the SYK model, arXiv:1711.08482.
%\bibitem{Das2}
%S. R. Das, A. Jevicki, and K. Suzuki, \emph{Three dimensional view of the SYK/AdS duality}, J. High Energ. Phys. 2017, 17 (2017),	arXiv:1704.07208.
\bibitem{Stanford}
J. Maldacena and D. Stanford, \emph{Remarks on the Sachdev-Ye-Kitaev model}, Phys. Rev. D {\bf94}, 106002 (2016), arXiv:1604.07818.
\bibitem{Polchinski}
J. Polchinski and V. Rosenhaus, \emph{The Spectrum in the Sachdev-Ye-Kitaev Model}, JHEP {\bf1604}, 001 (2016), arXiv:1601.06768.
%\cite{Maldacena:2015waa}
%1461 citations counted in INSPIRE as of 30 Jan 2023
\bibitem{csyk1}
S. Cao, Y. C. Rui and X. H. Ge, \emph{Thermodynamic phase structure of complex Sachdev-Ye-Kitaev model and charged black hole in deformed JT gravity}, arXiv:2103.16270.
\bibitem{csyk2}
J. Louw and S. Kehrein, \emph{The shared universality of charged black holes and the many many-body SYK model}, arXiv: 2204.09629.
\bibitem{Godet}
V. Godet and C. Marteau, \emph{New boundary conditions for $AdS_2$}, arXiv:2005.08999.
\bibitem{Ahn:2020bks}
Y.~Ahn, V.~Jahnke, H.~S.~Jeong, K.~Y.~Kim, K.~S.~Lee, and M.~Nishida,\emph{Pole-skipping of scalar and vector fields in hyperbolic space: conformal blocks and holography}, JHEP \textbf{09}, 111 (2020), arXiv:2006.00974.
\bibitem{Louis-Martinez}
D. Louis-Martinez and G. Kunstatter, \emph{On Birckhoff's theorem in 2-D dilaton gravity}, Phys. Rev. D {\bf49}, 5227 (1994).
\bibitem{Achucarro}
A. Achucarro and M. E. Ortiz, \emph{Relating black holes in two-dimensions and three-dimensions}, Phys. Rev. D {\bf48}, 3600 (1993), arXiv:hep-th/9304068.
\bibitem{Gegenberg}
J. Gegenberg, G. Kunstatter, and D. Louis-Martinez, \emph{Observables for two-dimensional black holes}, Phys. Rev. D {\bf51}, 1781 (1995), arXiv:gr-qc/9408015.
\bibitem{Gross1}
D. J. Gross and V. Rosenhaus, \emph{A generalization of Sachdev-Ye-Kitaev}, J. High Energ. Phys. {\bf2017}, 93 (2017), arXiv:1610.01569.
\bibitem{Kitaev2}
A. Kitaev and S.J. Suh, \emph{The soft mode in the Sachdev-Ye-Kitaev model and its gravity dual}, J. High Energ. Phys. {\bf2018}, 183 (2018),	arXiv:1711.08467.
\bibitem{Rosenhaus}
V. Rosenhaus, \emph{An introduction to the SYK model}, J. Phys. A: Math. Theor. {\bf52} 323001 (2019), arXiv:1807.03334.
\bibitem{Maria}
M. Tikhanovskaya, H. Guo, S. Sachdev, and G. Tarnopolsky, \emph{Excitation spectra of quantum matter without quasiparticles. I. Sachdev-Ye-Kitaev models}, Phys. Rev. B {\bf103}, 075141 (2021), arXiv:2010.09742.
\bibitem{Gross3}
D. J. Gross and V. Rosenhaus, \emph{A line of CFTs: from generalized free fields to SYK}, J. High Energ. Phys. 2017, 86 (2017), arXiv:1706.07015.
\bibitem{Gross2}
D. J. Gross and V. Rosenhaus, \emph{All point correlation functions in SYK}, J. High Energ. Phys. 2017, 148 (2017), arXiv:1710.08113.
\bibitem{Thomas}
T. Faulkner, H. Liu, J. McGreevy, and D. Vegh, \emph{Emergent quantum criticality, Fermi surfaces, and AdS$_2$}, Phys. Rev. D 83, 125002, arXiv:0907.2694.
\bibitem{Faulkner}
T. Faulkner, N. Iqbal, H. Liu, J. McGreevy, and D. Vegh, \emph{Holographic non-Fermi-liquid fixed points}, Phil. Trans. R. Soc. A {\bf369}, 1640 (2011), arXiv:1101.0597.
\bibitem{Gu}
Y. Gu, A. Kitaev, S. Sachdev, and G. Tarnopolsky, \emph{Notes on the complex Sachdev-Ye-Kitaev model}, JHEP {\bf2020}, 157 (2020), arXiv:1910.14099.
\bibitem{Subir}
S. Sachdev, \emph{Bekenstein-Hawking Entropy and Strange Metals}, Phys. Rev. X {\bf5}, 041025 (2015), arXiv:1506.05111.
\bibitem{Sen1}
A. Sen, \emph{Black Hole Entropy Function and the Attractor Mechanism in Higher Derivative Gravity}, JHEP {\bf09} 038 (2005), arXiv:hep-th/0506177.
\bibitem{Sen2}
A. Sen, \emph{Entropy Function and AdS(2)/CFT(1) Correspondence}, JHEP {\bf11} 075 (2008), arXiv:0805.0095.
\bibitem{Parcollet}
O. Parcollet, A. Georges, G. Kotliar, and A. Sengupta, \emph{Overscreened Multichannel SU(N) Kondo Model: Large-N Solution and Conformal Field Theory}, Phys. Rev. B 58, 3794 (1998), arXiv:cond-mat/9711192.
\bibitem{Georges}
A. Georges, O. Parcollet, and S. Sachdev, \emph{Quantum Fluctuations of a Nearly Critical Heisenberg Spin Glass}, Phys. Rev. B 63, 134406 (2001), arXiv:cond-mat/0009388.
\bibitem{Cai}
R. G. Cai, Y. H. Qi, Y. L. Wu, and Y. L. Zhang, \emph{Topological non-Fermi liquid}, Phys. Rev. D {\bf95}, 124026 (2017), arXiv:1601.03865.
\bibitem{Wilczek}
F. Wilczek and A. Zee, \emph{Families from spinors}, Phys. Rev. D {\bf25}, 553 (1982).

%\bibitem{Ahn:2020baf}
%Y.~Ahn, V.~Jahnke, H.~S.~Jeong, K.~S.~Lee, %M.~Nishida and K.~Y.~Kim,
%JHEP \textbf{03} (2021), 175
%doi:10.1007/JHEP03(2021)175
%[arXiv:2010.16166 [hep-th]].


% Please avoid comments such as "For a review'', "For some examples",
% "and references therein" or move them in the text. In general,
% please leave only references in the bibliography and move all
% accessory text in footnotes.

% Also, please have only one work for each \bibitem.


\end{thebibliography}
\end{document}